\documentclass[useAMS,usenatbib]{mn2e}

\usepackage{graphicx}
\usepackage{verbatim}
\usepackage{fix2col}

\title[Environments of SDSS galaxy classes]{The nature of the Sloan Digital Sky Survey galaxies in various classes based on morphology, colour and spectral features -- III. Environments}
\author[J. H. Lee et al.]{Joon Hyeop Lee$^{1,2}$\thanks{E-mail:
jhlee@astro.snu.ac.kr (JHL); mglee@astrog.snu.ac.kr (MGL); cbp@kias.re.kr (CBP); yychoi@kias.re.kr (YYC)}, Myung Gyoon Lee$^{2\star}$, Changbom Park$^{3\star}$, Yun-Young Choi$^{4\star}$\\
$^{1}$Korea Astronomy and Space Science Institute, Daejeon 305-348, Korea\\
$^{2}$Astronomy Program, Department of Physics and Astronomy, Seoul National University, Seoul 151-742, Korea\\
$^{3}$Korea Institute for Advanced Study, Dongdaemun-gu, Seoul 103-722, Korea\\
$^{4}$Astrophysical Research Centre for the Structure and Evolution of the Cosmos, Sejong University, Seoul 143-747, Korea}
\begin{document}

\date{Accepted 2009 November 23. Received 2009 November 23; in original form 2009 September 09}

\pagerange{\pageref{firstpage}--\pageref{lastpage}} \pubyear{2009}

\maketitle

\label{firstpage}

\begin{abstract}
We present a study on the environments of the SDSS galaxies
divided into fine classes based on their morphology, colour and spectral features.
The SDSS galaxies are classified into early-type and late-type; red and blue;
passive, H{\protect\scriptsize II}, Seyfert and LINER, which returns a total of 16 fine classes of galaxies.
We estimate the local number density, target-excluded local luminosity density, local colour, close pair fraction and the luminosity and colour of the brightest neighbour, which are compared between the fine classes comprehensively.
The morphology-colour class of galaxies strongly depends on the local density, with the approximate order of high-density preference: red early-type galaxies (REGs) -- red late-type galaxies (RLGs) -- blue early-type galaxies (BEGs) -- blue late-type galaxies (BLGs).
We find that high-density environments (like cluster environments) seem to suppress AGN activity.
The pair fraction of H{\protect\scriptsize II} REGs does not show statistically significant difference from that of passive REGs, while the pair fraction of H{\protect\scriptsize II} BLGs is smaller than that of non-H{\protect\scriptsize II} BLGs.
H{\protect\scriptsize II} BLGs show obvious double (red + blue) peaks in the distribution of the brightest neighbour colour, while red galaxies show a single red peak.
The brightest neighbours of Seyfert BLGs tend to be blue, while those of LINER BLGs tend to be red, which implies that the difference between Seyfert and LINER may be related to the pair interaction.
Other various environments of the fine classes are investigated, and their implication on galaxy evolution is discussed.
\end{abstract}


\section{Introduction}

Galaxies in the Universe have a wide range of physical properties such as morphology, spectra, mass, and so on. One of the major goals of the extragalactic astronomy is to understand what determines such properties of galaxies, and the first step toward the goal is to understand exactly how different the various galaxies are.
One efficient way to comprehend the nature of the diverse galaxies is to classify them using some criteria and to compare their statistical properties between the classes.
In many previous studies, however, galaxies have been classified using simple schemes based on just one or two properties, which sometimes miss detailed aspects of galaxy evolution. For example, several unusual classes of galaxies such as blue early-type galaxies \citep{fer05,lee06} and passive spiral galaxies \citep{yam04,ish07}, or the fundamental parameter difference between finely-divided classes in the same morphology class \citep{lee07,cho09a} are difficult to investigate using simple classification schemes.
Recently, \citet[][hereafter Paper I and Paper II, respectively]{lee08,lee09} presented studies on the optical and multi-wavelength properties of the Sloan Digital Sky Survey \citep[SDSS;][]{yor00} galaxies divided into fine classes based on their morphology, colour and spectral features.
Those finely-classified SDSS galaxies show distinguishable differences from each other in their photometric (both optical and multi-wavelength) and structural properties, which are hardly found in the studies using simple classification schemes.
Based on the understanding of the basic properties of the various galaxy classes presented in Paper I and Paper II, now it is needed to approach the next question: what makes them have such diverse natures?

For a long time, the environments of galaxies have been regarded as important factors determining the nature of galaxies.
Since close galaxies have influence on each other directly or indirectly, the local density about a galaxy is closely related to the properties of that galaxy \citep{kue05,par07}. 
Environments in which galaxies have many chances to interact with each other, induce morphology transformation by the tidal interaction or merging of galaxies \citep{spi51,mer83,coz07}. Such galaxy interactions convert the properties of galaxies, by triggering star formation \citep{woo06} or by suppressing star formation with tidal gas stripping \citep{mar03}. Moreover, the complete merging of two or more galaxies is a classical candidate of the early-type galaxy formation mechanism \citep{too77,sea78}.
In addition to the direct interaction between galaxies, dense environments affect galaxies with their strong gravitational potential \citep{bek99}.
Since large amount of hot gas is concentrated in high-density environments \citep[e.g. galaxy clusters;][]{chu96}, the morphology transformation of galaxies happens in such environments by the \emph{ram pressure stripping} \citep{gun72,qui00,roe05}.

Due to the close relationships between environments and galaxy properties, many studies have been focused on the environmental effects in galaxy evolution.
\citet{bla05} inspected the environmental dependence of optical properties of the SDSS galaxies, finding that the colour is the most sensitive parameter to the local density, and that the other parameters are not sensitive to environments when the colour and luminosity are fixed.
\citet{kue05} found that the surface brightness profile type and axis ratio of galaxies are frequently correlated with the local density.
\citet{ber06} showed that the age, metallicity and $\alpha$-element enhancement of galaxies are affected by the environments of galaxies.
\citet{mat07} argued that galaxy evolution is accelerated in dense environments, which took place mainly at high redshifts, based on the age -- density relations with respect to galaxy luminosity or mass.
\citet{rog07} investigated the star formation history of early-type galaxies, showing that a significant number of galaxies in high-density environments undergo low but detectable recent star formation.
\citet{owe07} inspected the environmental dependence of starburst galaxies, finding that the starburst in a galaxy is triggered when its nearby (within 20 kpc) neighbour has similar luminosity or mass.
\citet{wes07} presented that most AGN host galaxies are closer to the red sequence than the blue sequence, but they prefer lower density environments than red sequence galaxies.
\citet{par07} inspected the environmental dependence of physical properties of the SDSS galaxies, based on the local density estimated using an adaptive smoothing kernel, showing that variations of galaxy properties are almost entirely due to the environmental dependence of morphology and luminosity, and that other properties are almost independent of local density, when morphology and luminosity are fixed.
\citet{van08} found that the role of satellite galaxies in the transformation from blue sequence galaxies to red sequence galaxies is important, in the sense that the satellite galaxies may quench the star formation of their host galaxies.
\citet{par08} and \citet{par09a} showed that the morphology of a galaxy depends sensitively on the distance and morphology of its nearest neighbour galaxy, and that galaxy properties in the general field do not directly depend much on the large-scale background density. They showed that the apparent dependence of galaxy properties on the background density environment can be largely explained by an accumulated result of the interactions between the nearest neighbour galaxies.

Since it is obvious that environments affect galaxy properties significantly, environments must have played an important role in determining the fine classes (i.e. morphology, colour, and spectral features) of galaxies as well. Thus, to understand the origin of the various fine galaxy classes, it is needed to compare their environments.
This is the third in the series papers on the nature of galaxies in various classes based on the morphology, colour and spectral features. In this paper, we investigate the environments of the finely-divided galaxy classes, using several environmental parameters.
The outline of this paper is as follows.
Section 2 shows the data set we used, and \S3 describes our classification scheme of the SDSS galaxies and the definition of the environmental parameters.
In \S4, we present the results, and their implication on galaxy evolution are discussed in \S5. The conclusions in this study are summarised in \S6.
Throughout this paper, we adopt the cosmological parameters 
$h=0.7$, $\Omega_{\Lambda}=0.7$, and $\Omega_{M}=0.3$.

\section{Data and Selection Criteria}

We briefly summarise the datasets, classification scheme and sample selection in this paper.
More details are described in Paper I and Paper II.

\subsection{Sloan Digital Sky Survey}

We use the SDSS Data Release 4 \citep[DR4;][]{ade06} main galaxy spectroscopic sample.
The SDSS is a photometric and spectroscopic survey, mapping about one quarter of the whole sky, and the SDSS DR4 spectroscopic data cover 4783 deg$^{2}$. The median FWHM of point sources in the $r$ band images is $1.4''$, and the wavelength coverage in the spectroscopy is 3800 -- 9200{\AA}.
We use both the SDSS pipeline \citep{sto02} data and the Max-Planck-Institute for Astronomy catalogue \citep[MPA catalogue;][]{kau03a,tre04,gal06}.
Using the SDSS atlas images, we have measured the colour gradient, inverse concentration index, and the axis ratio of galaxies corrected for the inclination and the seeing effects \citep{cho07}.
Corrections of the SDSS magnitudes have been conducted for the foreground extinction \citep{sch98}, the redshift effect \citep[K-correction;][]{bla03} and galaxy luminosity evolution \citep[evolutionary correction;][]{teg04}. We corrected the observed magnitudes of about 360,000 SDSS galaxies into the magnitudes at redshift z = 0.1, because the SDSS galaxies were observed most frequently there.
Since the corrections are optional in this paper, we denote the magnitude with K-correction as $^{0.1{\textrm{\protect\scriptsize K}}}m$ and the magnitude with both K-correction and evolutionary correction as $^{0.1{\textrm{\protect\scriptsize KE}}}m$, if the observed magnitude is $m$.

\subsection{Additional multi-wavelength datasets}

As a near-infrared dataset, we use the Two Micron All Sky Survey \citep[2MASS;][]{jar00} Extended Source Catalogue (XSC). The number of 2MASS extended sources matched with our SDSS sample is about 200,000, with false match probability of 0.005 per cent.
The InfraRed Astronomical Satellite \citep[IRAS;][]{neu84} Faint Source Catalogue \citep[FSC;][]{mos92} is used as a mid- and far-infrared dataset. About 7,000 IRAS sources were matched with our SDSS sample, and the false match probability between SDSS and IRAS is about 0.071 per cent.
We use the Faint Images of the Radio Sky at Twenty-centimetres \citep[FIRST;][]{bec95} data, and the number of the matched FIRST sources is about 14,000, with 0.45 per cent false match probability.
The Galaxy Evolution Explorer \citep[GALEX;][]{mar03} survey GR2/GR3 data are used as an ultraviolet dataset. About 85,000 GALEX objects were matched, with the false match probability about 1.46 per cent.
As an X-ray dataset, we use the Roentgen Satellite \citep[ROSAT;][]{asc81} all sky survey data. The information of the ROSAT sources matched with the SDSS objects provided by the SDSS is used, and the number of the ROSAT sources is about 2,200.

\begin{figure}
\includegraphics[width=84mm]{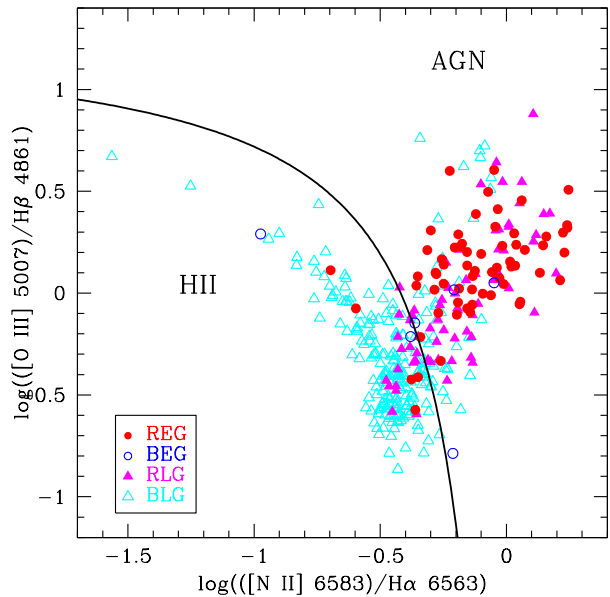}
\caption{ Morphology and colour class distribution on the BPT diagram for a small sample of the SDSS galaxies. Different symbols indicate different morphology-colour classes: REGs (filled circle), BEGs (open circle), RLGs (filled star) and BLGs (open star). The line shows the boundary dividing H{\protect\scriptsize II} galaxies and AGN host galaxies in this paper. }
\label{class}
\end{figure}

\begin{figure}
\includegraphics[width=84mm]{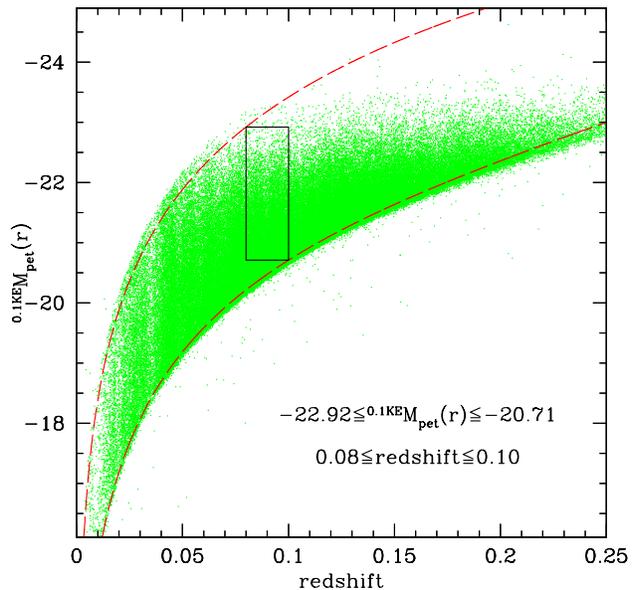}
\caption{ Absolute magnitude versus redshift for the SDSS galaxies. The solid-line box represents the sample volume selected in this paper. The dashed lines are the limits of the complete spectroscopic sample.}
\label{volumes}
\end{figure}

\begin{table}
\centering
\caption{Abbreviations of the 16 fine galaxy classes}
\label{tclcl}
\begin{tabular}{lcccc}
\hline \hline
& \multicolumn{2}{c}{Early-type} & \multicolumn{2}{c}{Late-type} \\
& Red & Blue & Red & Blue \\
\hline
Passive & $p$REG & $p$BEG & $p$RLG & $p$BLG \\
H{\protect\scriptsize II} & $h$REG & $h$BEG & $h$RLG & $h$BLG \\
Seyfert & $s$REG & $s$BEG & $s$RLG & $s$BLG \\
LINER & $l$REG & $l$BEG & $l$RLG & $l$BLG \\
\hline \hline
\end{tabular}
\end{table}

\begin{table}
\centering
\caption{The number of galaxies in each fine class}
\label{clnum}
\begin{tabular}{lrrrrr}
\hline \hline
& REG & BEG & RLG & BLG & Total\\
\hline
Passive & 5109 & 26 & 446 & 77	&	5658\\
H{\protect\scriptsize II} & 2479 & 148 & 1626 & 9632 &	13885\\
Seyfert & 1477 & 98 & 1770 & 1472 &	4817\\
LINER & 4829 & 159 & 3569 & 1281 &	9838\\
\hline
Total & 13894 & 431 & 7411 & 12462 &	34198 \\
\hline \hline
\end{tabular}
\end{table}

\subsection{Classifications and volume-limited sample}

In this paper, we classify the SDSS galaxies using three criteria: morphology, colour and spectral features.
Here, we briefly introduce the galaxy classification scheme. More details about classifications and possible biases are described in Paper I and Paper II.
First, the SDSS galaxies were classified into early-type galaxies and late-type galaxies in the colour -- colour gradient -- light-concentration parameter space \citep{par05}.
Second, we divided the SDSS galaxies into red galaxies and blue galaxies, based on the colour distribution of early-type galaxies as a function of redshift \citep{lee06}.
Third, the SDSS galaxies were classified into passive galaxies, H{\protect\scriptsize II} galaxies, Seyfert galaxies and low ionisation nuclear emission region (LINER) galaxies, based on their flux ratios between several spectral lines \citep[e.g. BPT diagram;][]{bal81,kau03b,kew06}.
Fig.~\ref{class} shows the distributions of different morphology and colour classes on the BPT diagram. As mentioned previously, it is noted that the three kinds of classification do not always agree with each other, showing the necessity of fine classification.
Those three galaxy classifications with different criteria return a total of 16 fine classes of galaxies: [early-type, late-type] $\times$ [red, blue] $\times$ [passive, H{\protect\scriptsize II}, Seyfert, LINER].
Hereafter, we use the following abbreviations for the 16 galaxy classes: REG (red early-type galaxy), BEG (blue early-type galaxy), RLG (red late-type galaxy), BLG (blue late-type galaxy), and $p$- (passive), $h$- (H{\protect\scriptsize II}), $s$- (Seyfert), $l$- (LINER). The abbreviations for the 16 fine galaxy classes are summarised in Table \ref{tclcl}.
For RLGs, we only use RLGs with their axis ratio larger than 0.6, because many BLGs with large inclinations may be classified into RLGs due to their internal dust extinction \citep{cho07}.

Since most properties of galaxies are known to be sensitive to their luminosity and redshift, the sample volume needs to be carefully controlled in statistical studies of galaxies.
Among the three volumes selected in Paper I and Paper II, we use the V1 sample as shown in Fig.~\ref{volumes}, which contains the largest number of galaxies and covers the brightest range of absolute magnitude among the three volumes: $-22.92 \le$ $^{0.1\textrm{\protect\scriptsize KE}}M_{pet}(r)$\footnote{The Petrosian absolute magnitude in the $r$ band, after K-correction and evolutionary correction.} $\le -20.71$ and $0.08\le$ z $\le0.10$.
The number of galaxies in each fine class in this sample is listed in Table \ref{clnum}, and more details about the sample-volume selection are described in Paper I.

\begin{figure*}
\includegraphics[width=80mm]{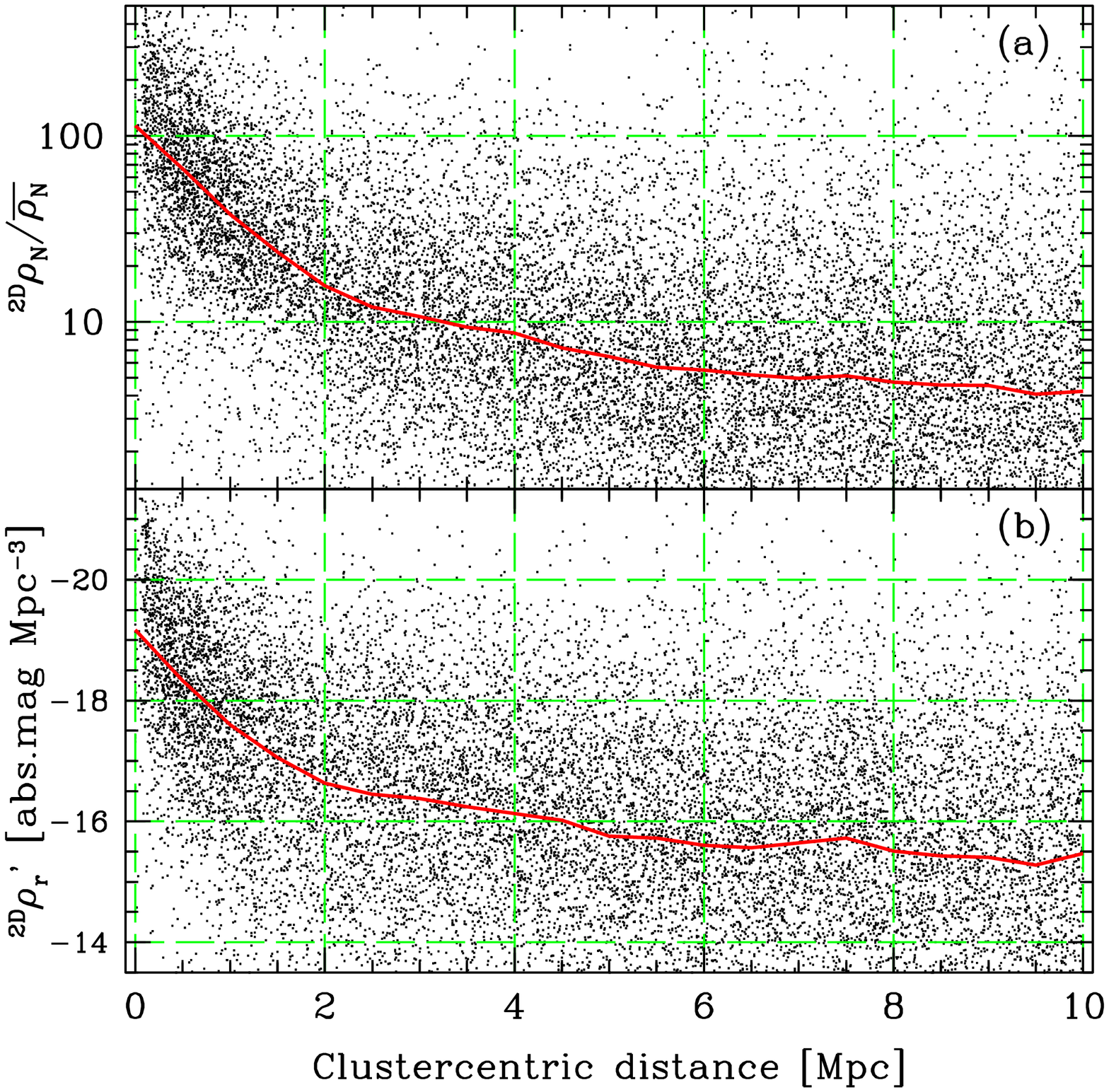}
\includegraphics[width=80mm]{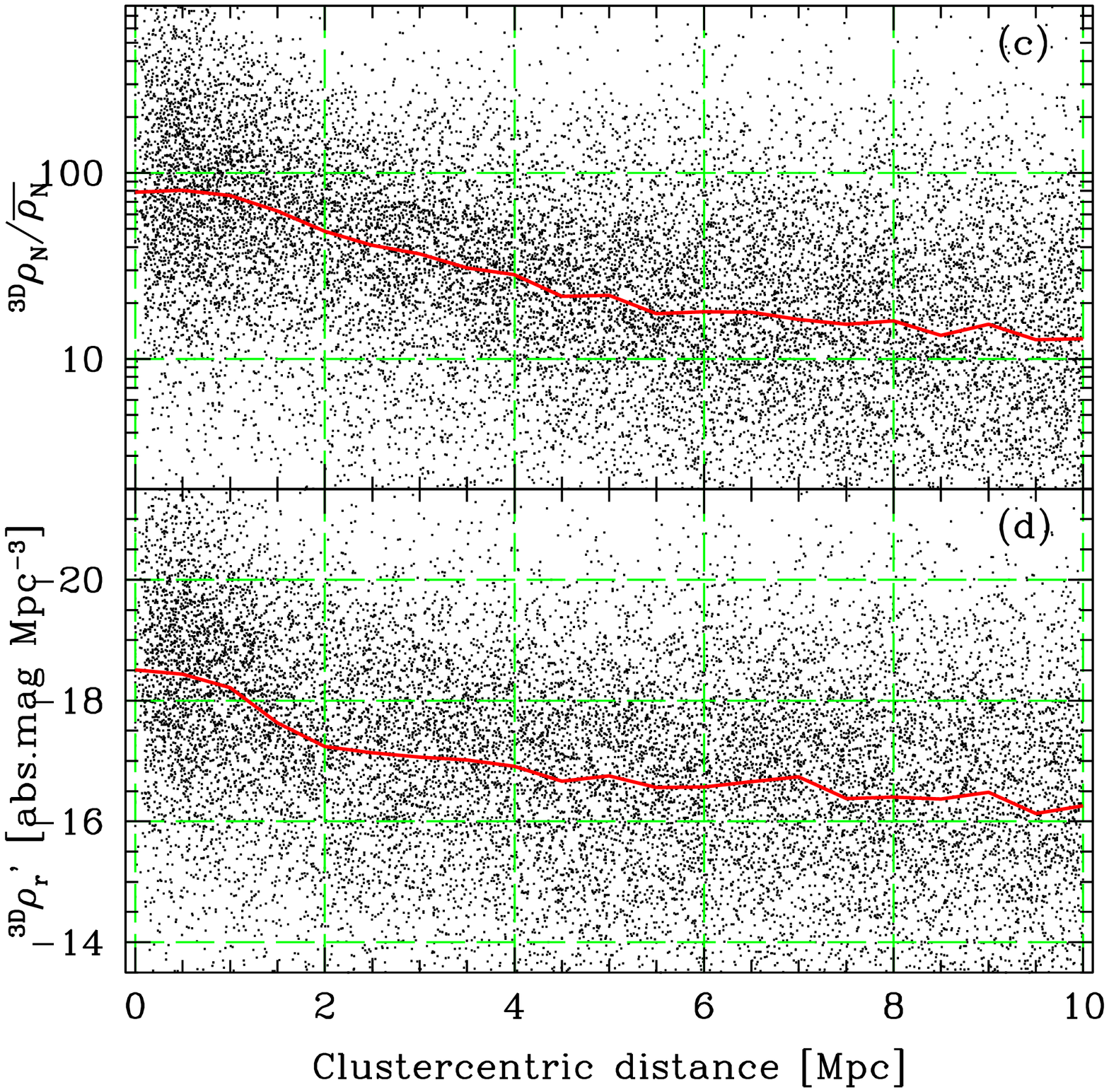}
\caption{ (a) $^{\textrm{\protect\tiny 2D}}{\rho_N}$ versus the clustercentric distance. (b) $^{\textrm{\protect\tiny 2D}}{\rho_r}'$ versus the clustercentric distance. (c) $^{\textrm{\protect\tiny 3D}}{\rho_N}$ versus the clustercentric distance. (d) $^{\textrm{\protect\tiny 3D}}{\rho_r}'$ versus the clustercentric distance. $\bar{\rho}_{N}$ ($=2.469\times10^{-3}$ Mpc$^{-3}$) is the mean number density in our sample. The clustercentric distance is defined as the projected comoving distance to the nearest galaxy cluster within the $\pm1000$ km s$^{-1}$ redshift slice. In each panel, the solid line connects the median local density values at given clustercentric distance bins.}
\label{cluster}
\end{figure*}

\begin{figure}
\includegraphics[width=84mm]{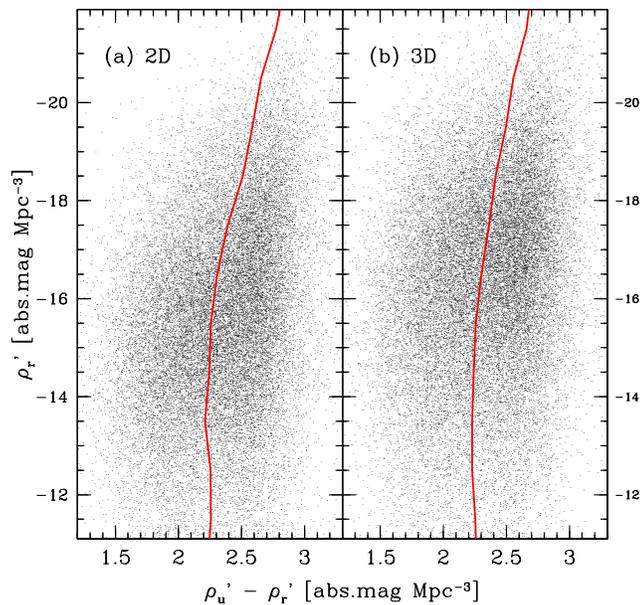}
\caption{ The target-excluded local luminosity density versus target-excluded local colour relation, based on (a) the 2D estimation and (b) the 3D estimation. The lines connect the median ${\rho_u}'-{\rho_r}'$ at given ${\rho_r}'$ bin. }
\label{allcmr}
\end{figure}

\section{Environmental Parameters}

\subsection{Local density indicators}\label{C2d3d}

There are two different ways in estimating the density of local environments, according to whether we define the local volume two-dimensionally (2D) or three-dimensionally (3D).
In the 2D method, the projected local density is estimated within a fixed redshift interval centred on the target galaxy. On the other hand, in the 3D method, the local density is estimated using the 3D distance calculated by regarding redshift difference purely as distance difference.
It is known that 2D-based parameters are more accurate in high-density environments, while 3D-based parameters are more accurate in low-density environments \citep{coo05}.
Since both methods have their own merits and demerits, we estimate both of them.
In this paper, we adopt $\pm1000$ km s$^{-1}$ as the redshift interval for the 2D local density estimation.
More detailed comparison between the 2D- and 3D-based environmental parameters are shown in \S\ref{A2d3d}.

The most frequently used parameter as a local density indicator is local number density ($\rho_N$). When estimating $\rho_N$, we adopt a spline smoothing kernel (see \S\ref{Akernel}). In addition, we use the comoving distance to the $n$-th nearest neighbour as the kernel size, which means that the smoothing kernel size is adaptive to the local number density.
The adaptive smoothing prevents oversmoothing in dense regions and preserves better the property of galaxy distributions than a smoothing with a fixed kernel size, as described by \citet{par07}, who used the spline-kernel with adaptive size enclosing a fixed number of the nearest neighbour galaxies.
In this paper, the local number density $\rho_N$ in Mpc$^{-3}$ is defined as:
\begin{equation}
\rho_N = \sum_{k=1}^n f_{sp}(d_k/d_n)/V
\end{equation}
where $n$ is an integer to define the kernel size $d_n$ as the comoving distance to the $n$-th nearest neighbour, and $d_k$ is the comoving distance to the $k$-th nearest neighbour. $f_{sp}(x)$ is the spline kernel function and $V$ is the local volume in Mpc$^{3}$.
It is noted that the $d_n$ and $d_k$ are \emph{projected} comoving distances in the 2D estimation. The local volume $V$ is a cylinder in the 2D estimation, the height of which is 2000 km s$^{-1}$, while it is a sphere in the 3D estimation.
From now on, we denote the 2D- and 3D-based local number densities as $^{\textrm{\protect\tiny{2D}}}\rho_N$ and $^{\textrm{\protect\tiny{3D}}}\rho_N$, respectively.

The local galaxy number density is the most frequently used local density indicator. However, the \emph{number} density does not reflect the local environments perfectly, because the properties of the counted galaxies are not considered in that parameter, such as their luminosity and colour. For this reason, local luminosity density or local mass density are also used as the local density indicators \citep{par08,par09a}.
In this paper, we estimate local luminosity density in the $r$ band ($\rho_r$), which is defined as:
\begin{equation}
\rho_r = -2.5 \log\bigg(\sum_{k=1}^n f_{sp}(d_k/d_n){\times}10^{-0.4\big(\,^{0.1\textrm{\protect\scriptsize K}}M_k(r)\big)}/V\bigg)
\end{equation}
in the unit of absolute-mag Mpc$^{-3}$, where $^{0.1\textrm{\protect\scriptsize K}}M_{k}(r)$ is the K-corrected $r$ band Petrosian absolute magnitude of the $k$-th nearest neighbour. In the same manner, the local luminosity density in the $u$ band ($\rho_u$) is defined, and the local colour is simply defined as $\rho_u-\rho_r$.
The behaviours of the local density indicators in this section are described in \S\ref{defld} with more details.
Throughout this paper, we adopt $n=5$ for all local density indicators, to focus on the small-scale environments of galaxies (see \S\ref{defld}).
We corrected the local luminosity densities by excluding the information of the target galaxy: target-excluded local luminosity density (${\rho_r}'$) and target-excluded local colour (${\rho_u}'-{\rho_r}'$). The necessity of this correction is described in \S\ref{Acorr}.

Since our sample luminosity range is biased to the very bright side of galaxy luminosity distribution, there is a caveat that the estimated local luminosity density and local colour may not exactly reflect the local environments. That is, the local luminosity density depends on the field-to-field variation of galaxy luminosity function. Moreover, the local colour may be bluer if we use a sample with more fainter luminosity range, because galaxy colours tend to be redder with increasing luminosity \citep{cho07}. This should be kept in mind throughout this paper.

Fig.~\ref{cluster} presents how ${\rho_N}$ and ${\rho_r}'$ are correlated with the clustercentric distance, based on the 2D and 3D estimation respectively. The clustercentric distance is defined as the projected comoving distance to the nearest known galaxy cluster\footnote{The coordinate list of the galaxy clusters was retrieved from the NASA Extragalactic Database (NED; http://nedwww.ipac.caltech.edu/).} within the $\pm1000$ km s$^{-1}$ redshift slice. At the clustercentric distance $<2$ Mpc, $^{\textrm{\protect\tiny 2D}}{\rho_r}'$ decreases rapidly as the clustercentric distance decreases, while the $^{\textrm{\protect\tiny 3D}}{\rho_r}'$ variation in the cluster regions is much smaller than the $^{\textrm{\protect\tiny 2D}}{\rho_r}'$ variation. If we adopt 1 -- 2 Mpc as the typical radius of galaxy clusters, the intermediate environments between galaxy clusters and fields approximately correspond to about $^{\textrm{\protect\tiny 2D}}{\rho_r}'\sim-17$ ($^{\textrm{\protect\tiny 2D}}{\rho_N}/\bar{\rho}_N\sim20$) and $^{\textrm{\protect\tiny 3D}}{\rho_r}'\sim-17.5$ ($^{\textrm{\protect\tiny 3D}}{\rho_N}/\bar{\rho}_N\sim60$), where $\bar{\rho}_{N}$ ($=2.469\times10^{-3}$ Mpc$^{-3}$) is the mean number density in our sample.
It is noted that the spectroscopic completeness in the central regions of rich clusters may be lower than that in the field \citep{yoo08}. This effect causes the underestimation of the local density in very-high-density environments, making the correlation between the local density and the clustercentric distance in Fig.~\ref{cluster} less clear.
Fig.~\ref{allcmr} shows the ${\rho_r}'$ versus ${\rho_u}'-{\rho_r}'$ plot, an `environmental colour-magnitude diagram'. Like the galaxy colour-magnitude diagram, the environmental colour-magnitude diagram shows a sequence, in the sense that brighter ${\rho_r}'$ environments have redder ${\rho_u}'-{\rho_r}'$.

\begin{figure*}
\includegraphics[width=80mm]{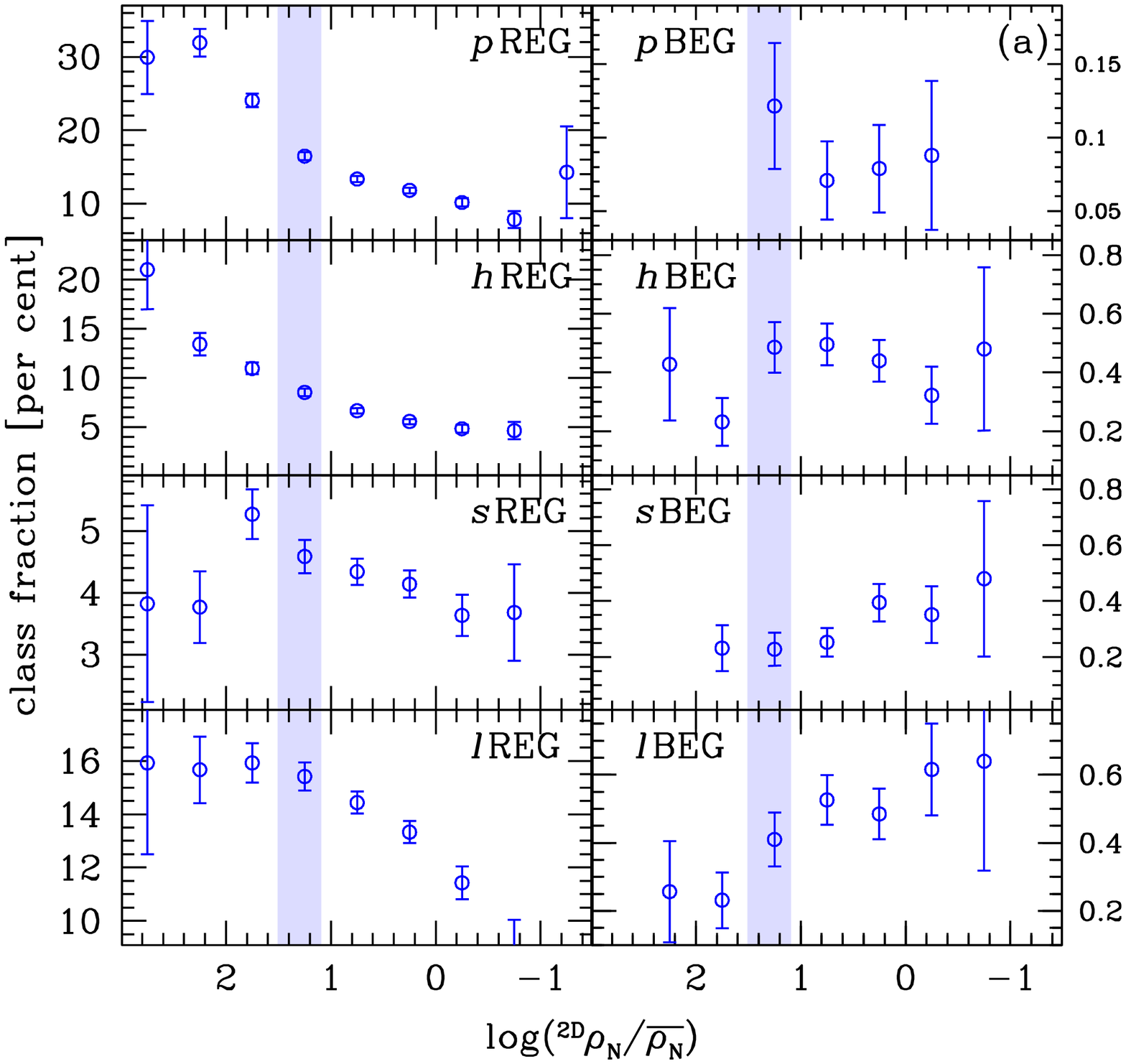}
\includegraphics[width=80mm]{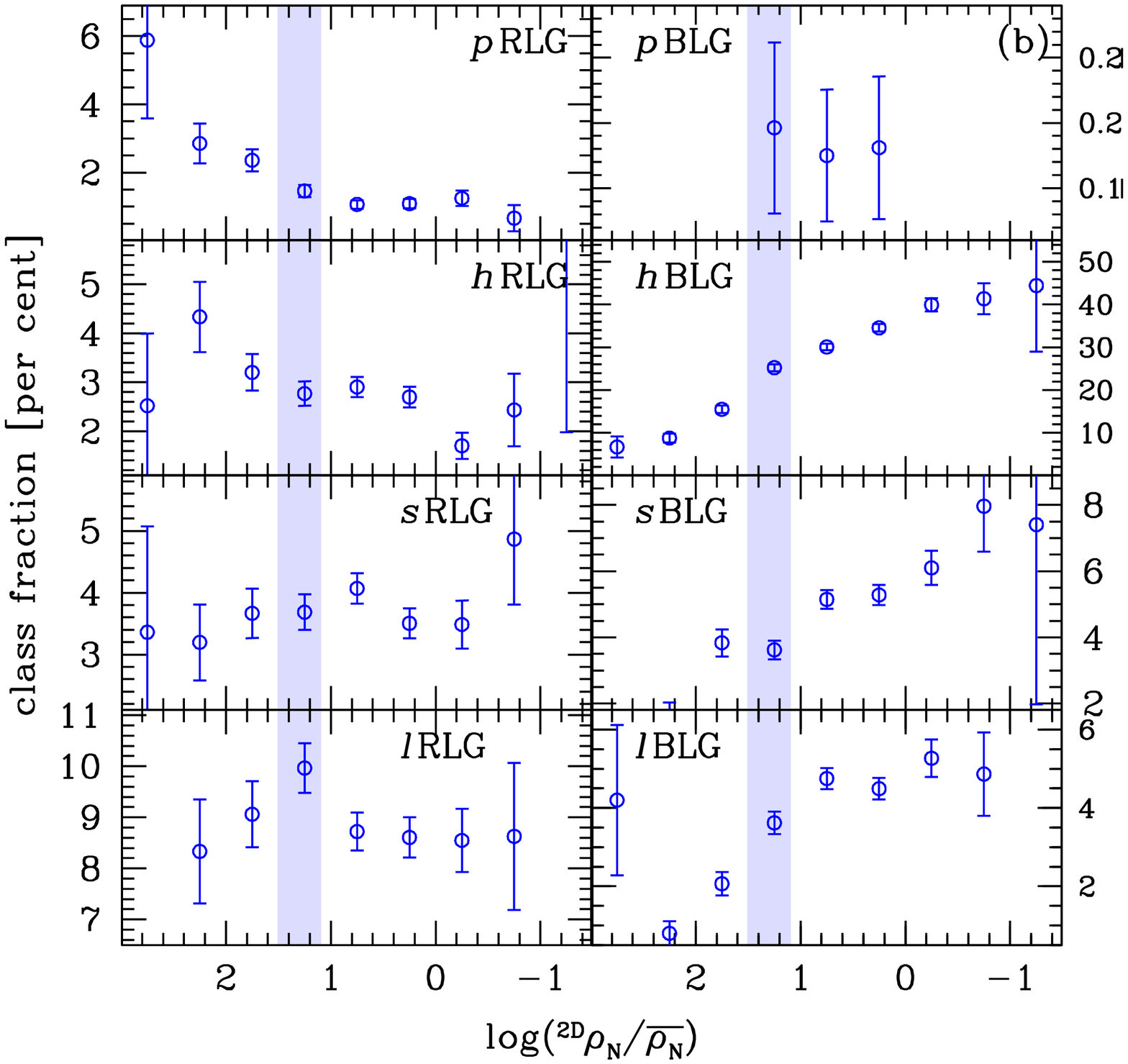}
\includegraphics[width=80mm]{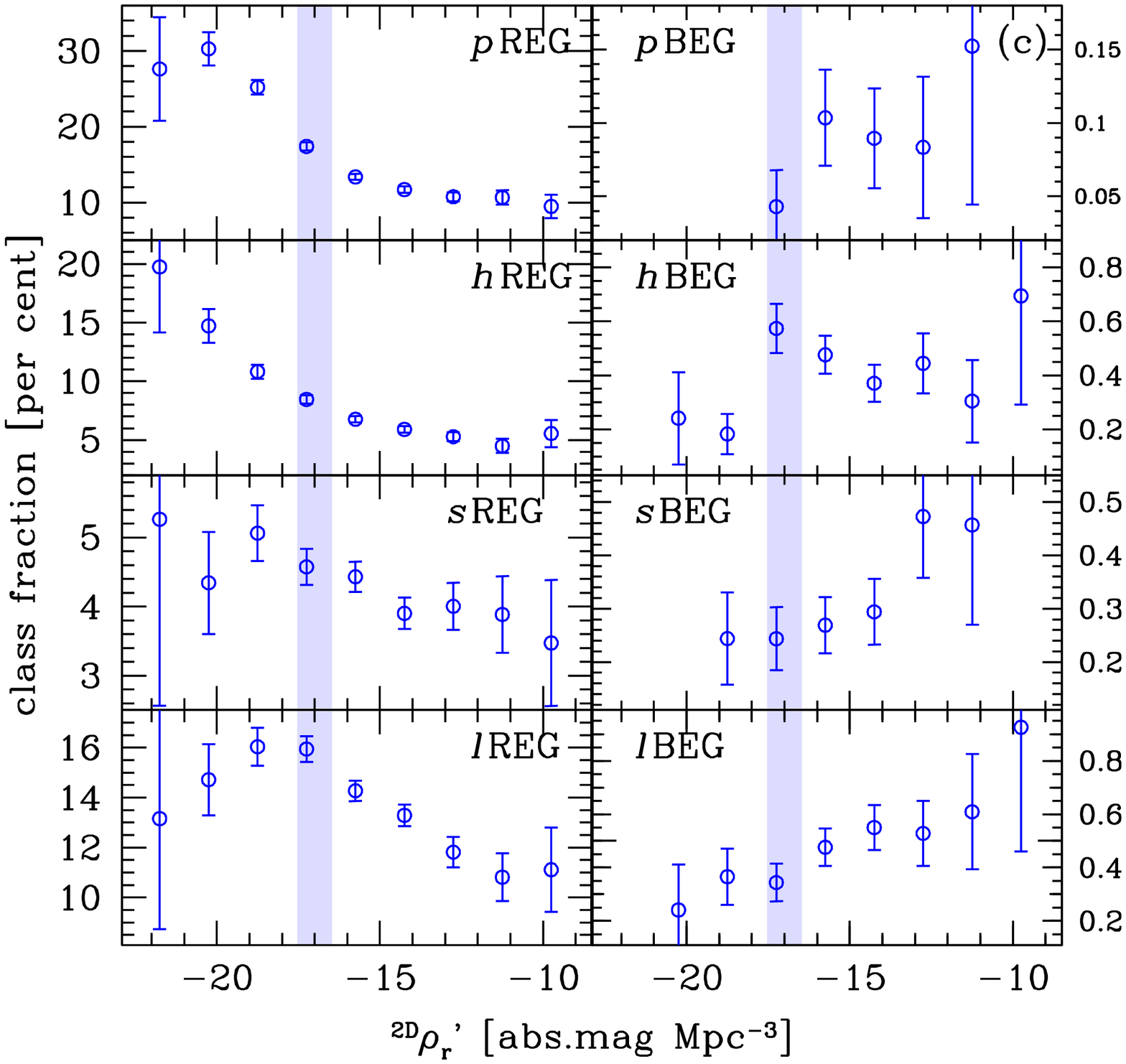}
\includegraphics[width=80mm]{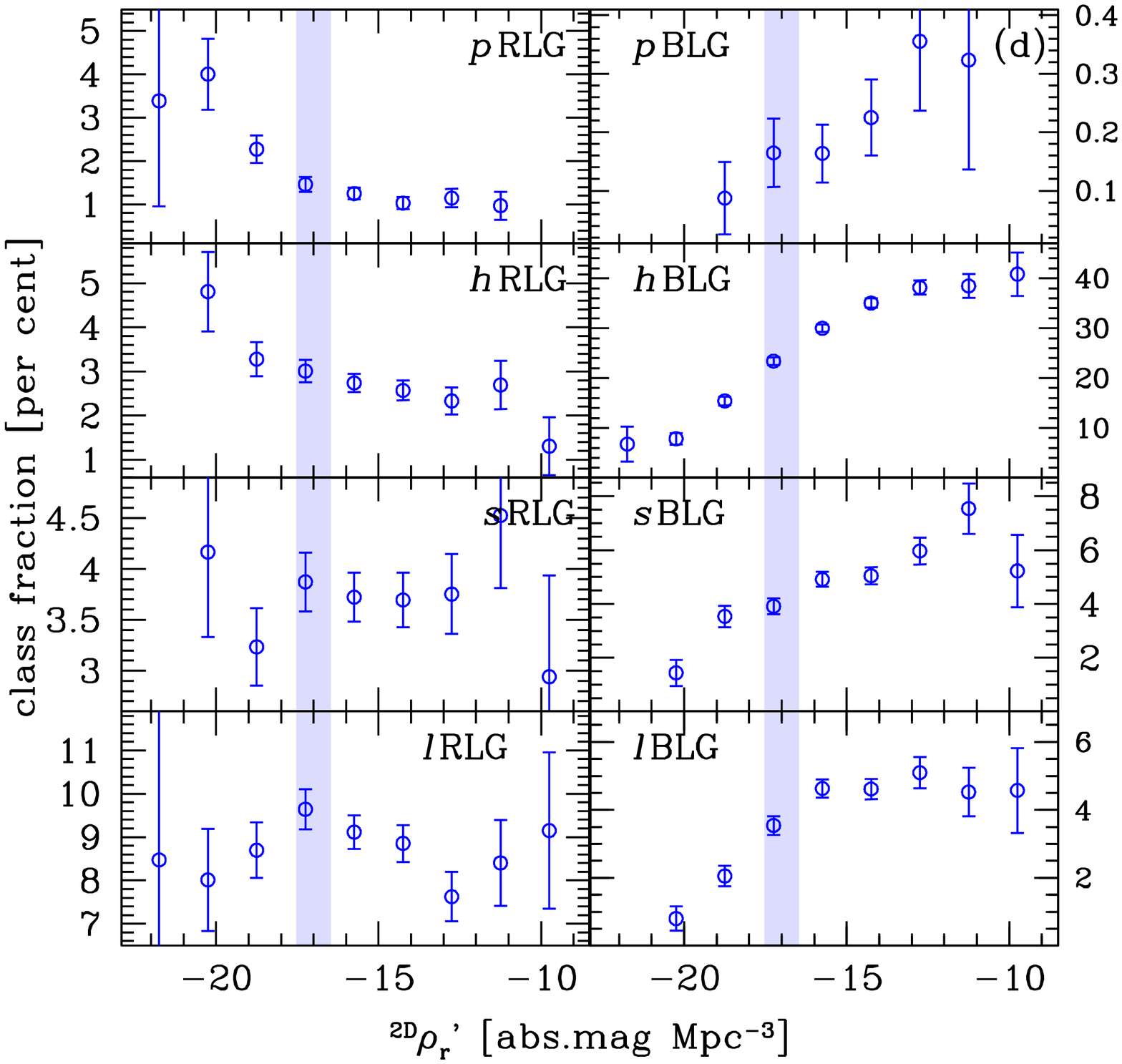}
\caption{ (a) The dependence of the early-type class fractions on the 2D local number density. (b) The same as (a), but for the late-type classes. (c) The dependence of the early-type class fractions on the target-excluded 2D local luminosity density. (d) The same as (c), but for the late-type classes.
Errorbars show the Poisson errors. The intermediate range between cluster and field environments is denoted as the shaded domain. }
\label{clfrac}
\end{figure*}

\subsection{Close pair fraction}

\citet{woo07} defined close pair galaxies as galaxies with their recession velocity difference $<500$ km s$^{-1}$ and their projected distance $<50$ kpc. Adopting the definition of \citet{woo07}, we estimate the pair fraction in each fine class.
We define the \emph{brightest neighbour} as the brightest one among the neighbour galaxies satisfying the close pair definition for a given target galaxy. The $r$-band absolute magnitude and $u-r$ colour of the brightest neighbour are denoted as $^{0.1{\textrm{\protect\scriptsize K}}}M(r)_{\textrm{\protect\scriptsize BrightestNeighbour}}$ and $^{0.1{\textrm{\protect\scriptsize K}}}(u-r)_{\textrm{\protect\scriptsize BrightestNeighbour}}$ respectively, the distribution of which will be compared between the fine classes, in the next section.
It is noted that the \emph{close pairs} in this paper do not necessarily mean that they are currently interacting pairs, which should be more strictly defined using the virial radii of the pair galaxies \citep{par08,par09a,par09b}. However, due to the practical difficulty in estimating the mass-to-light ratio of each fine class, we use this simple definition of close pairs, which selects pair galaxies that have relatively high chance of recent interactions.

\begin{figure*}
\includegraphics[width=80mm]{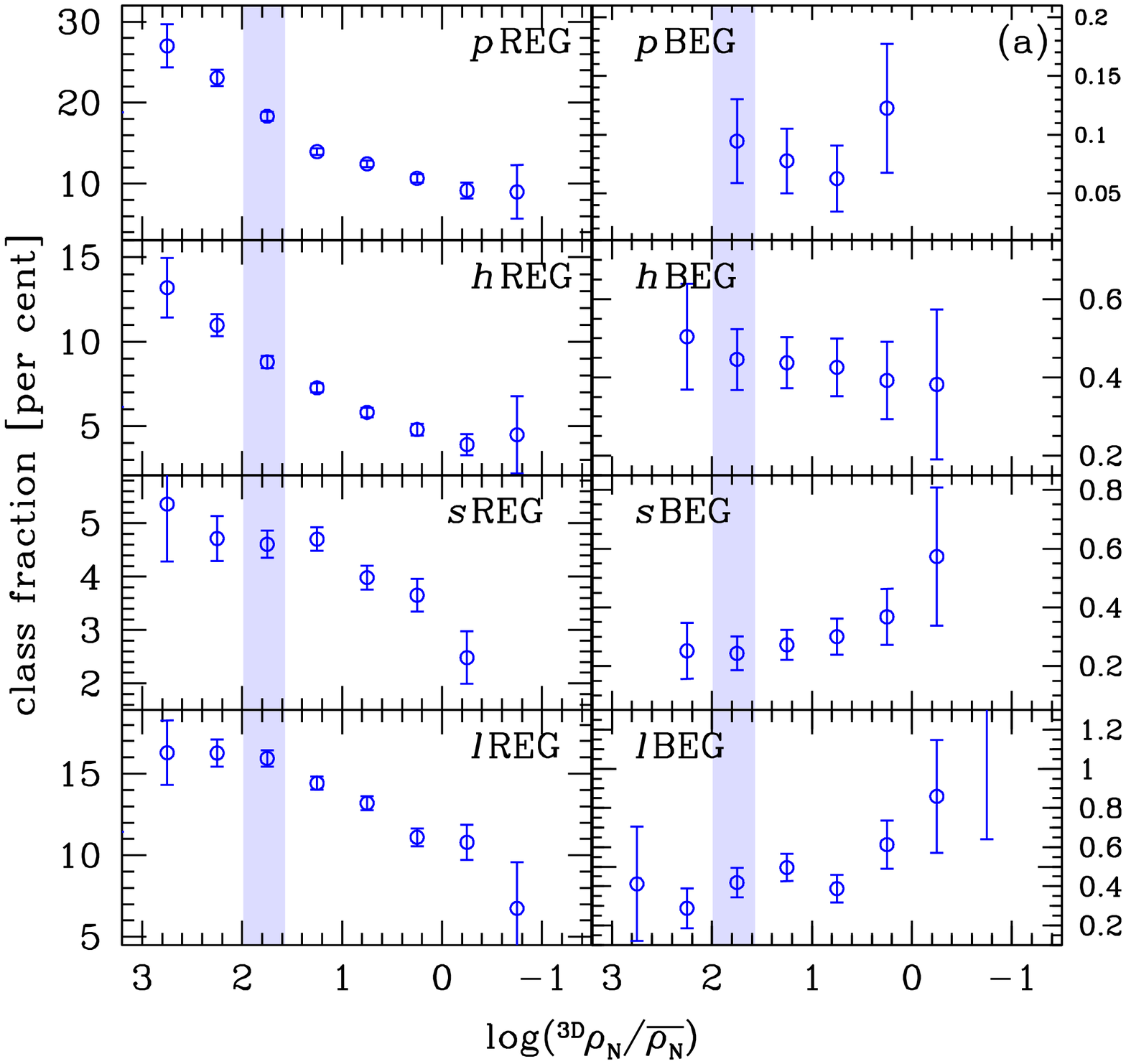}
\includegraphics[width=80mm]{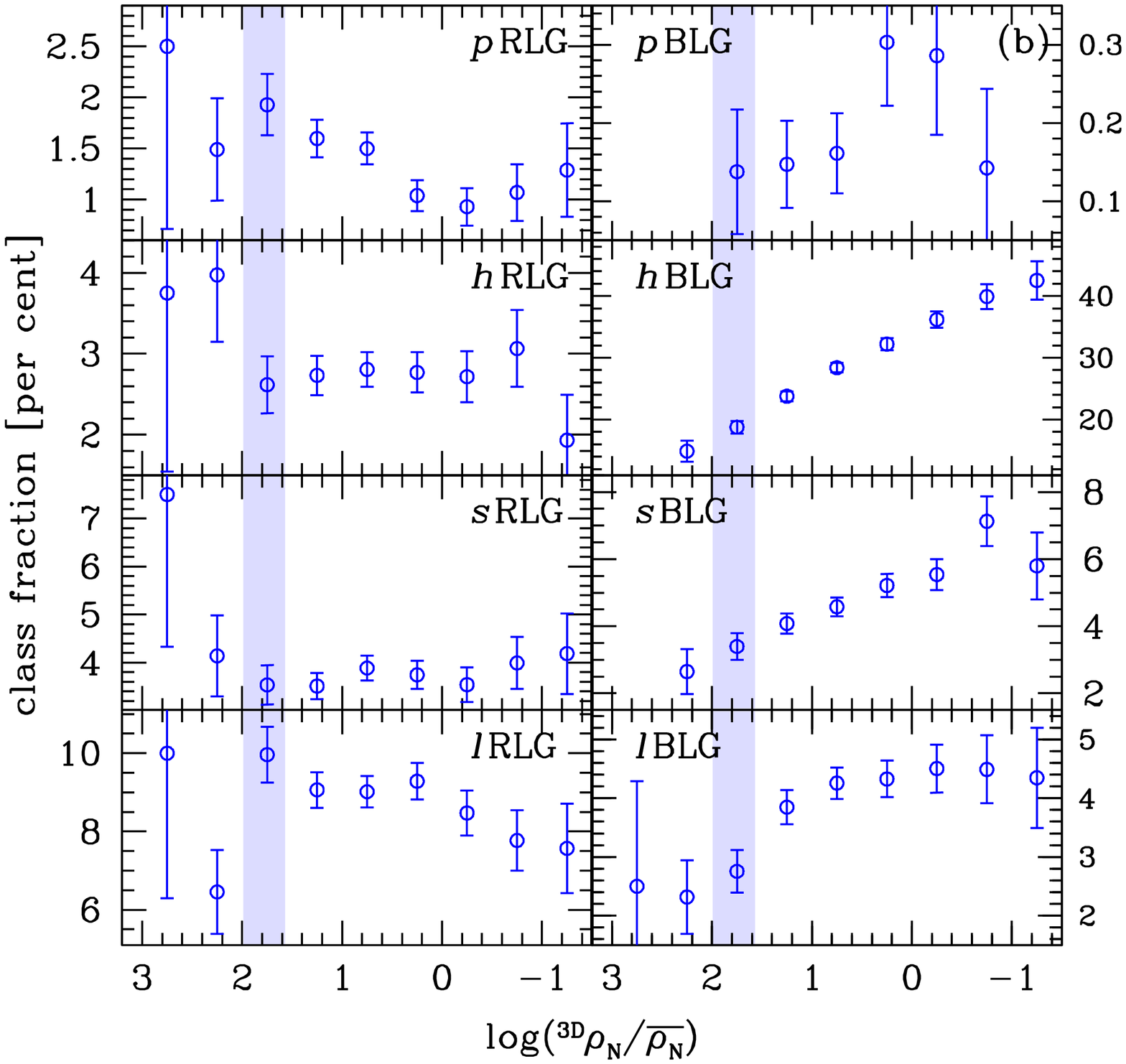}
\includegraphics[width=80mm]{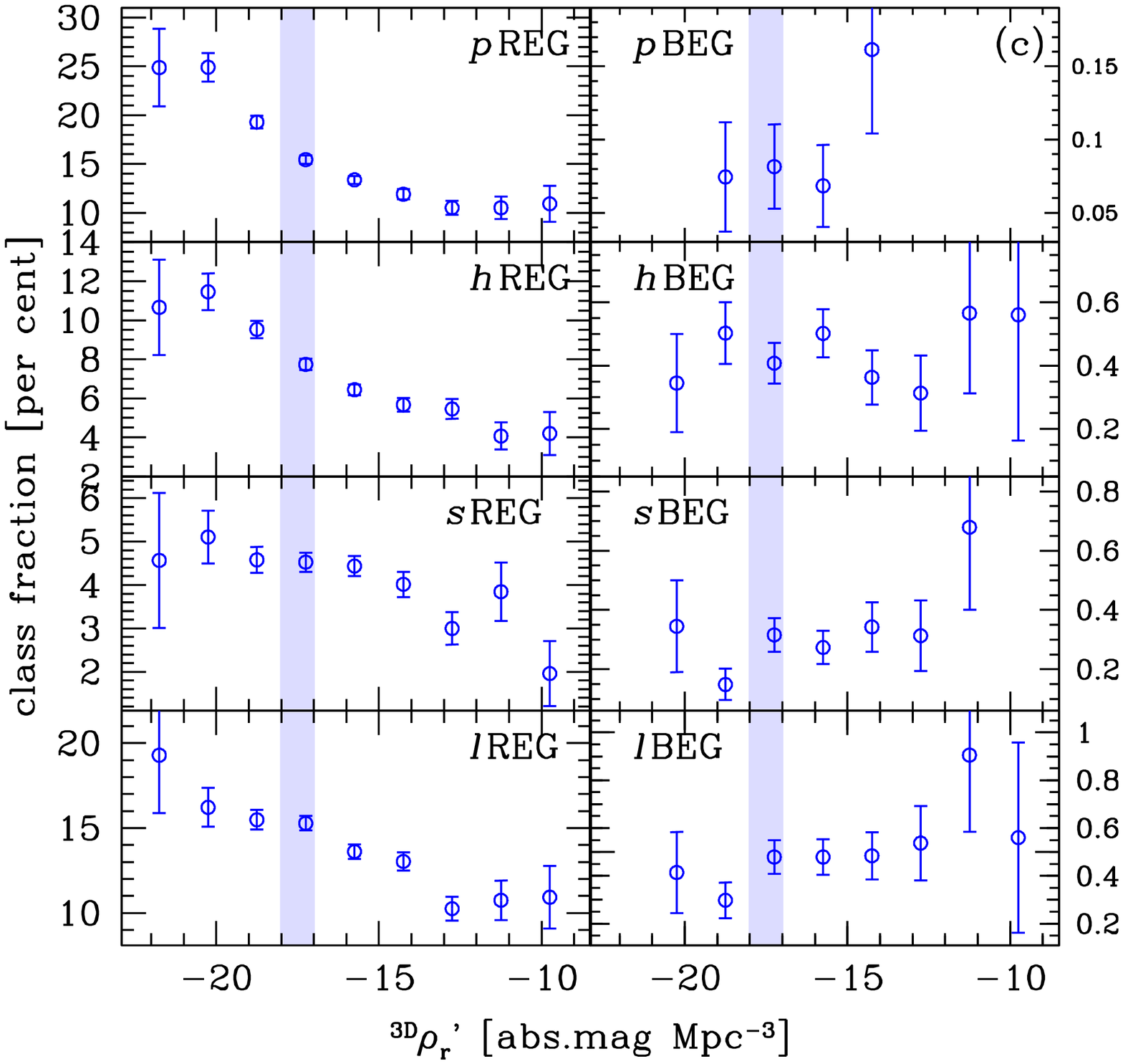}
\includegraphics[width=80mm]{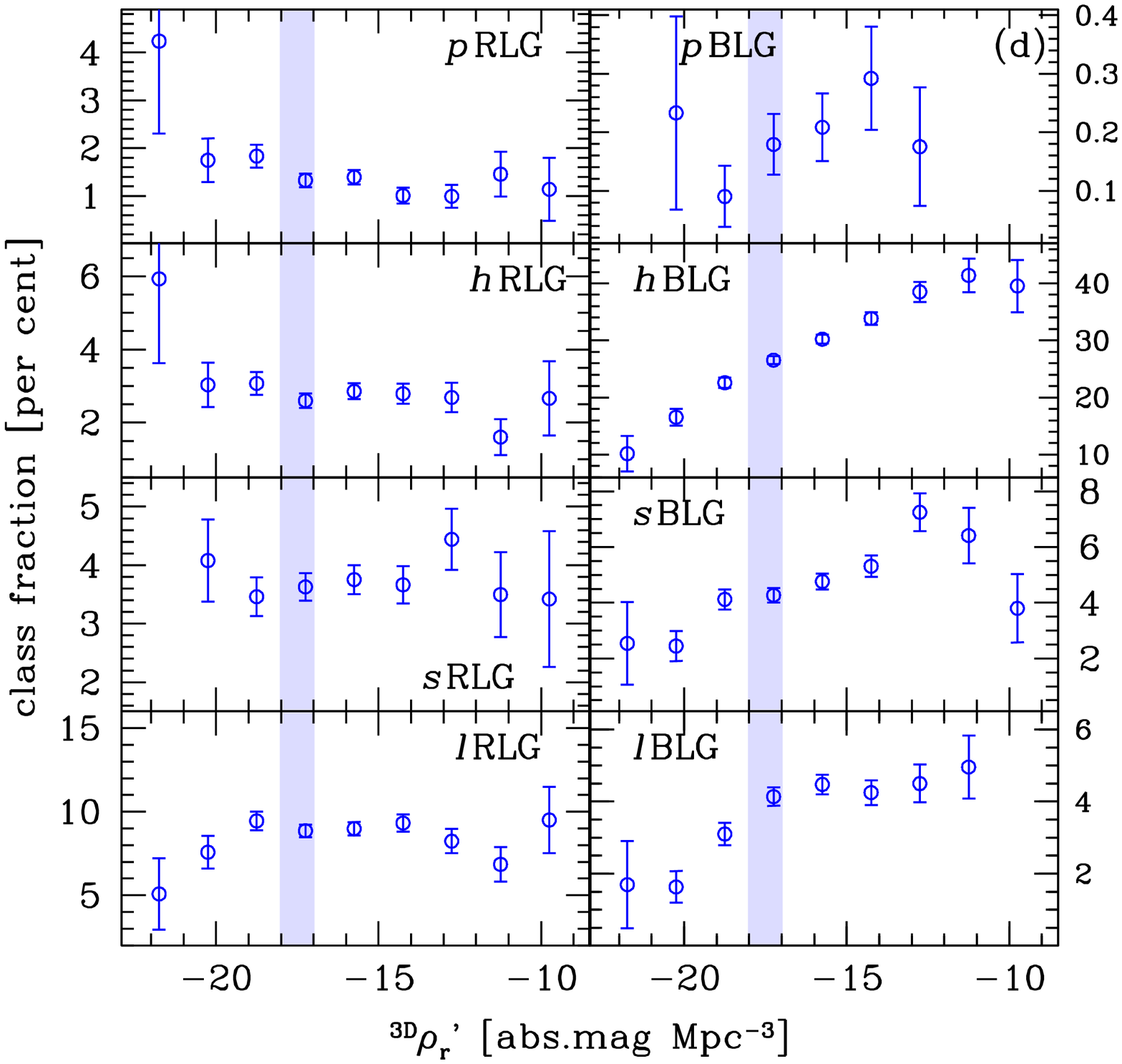}
\caption{ The same as Fig.~\ref{clfrac}, but for the 3D local density indicators. }
\label{clfrac3d}
\end{figure*}

\begin{figure*}
\includegraphics[width=168mm]{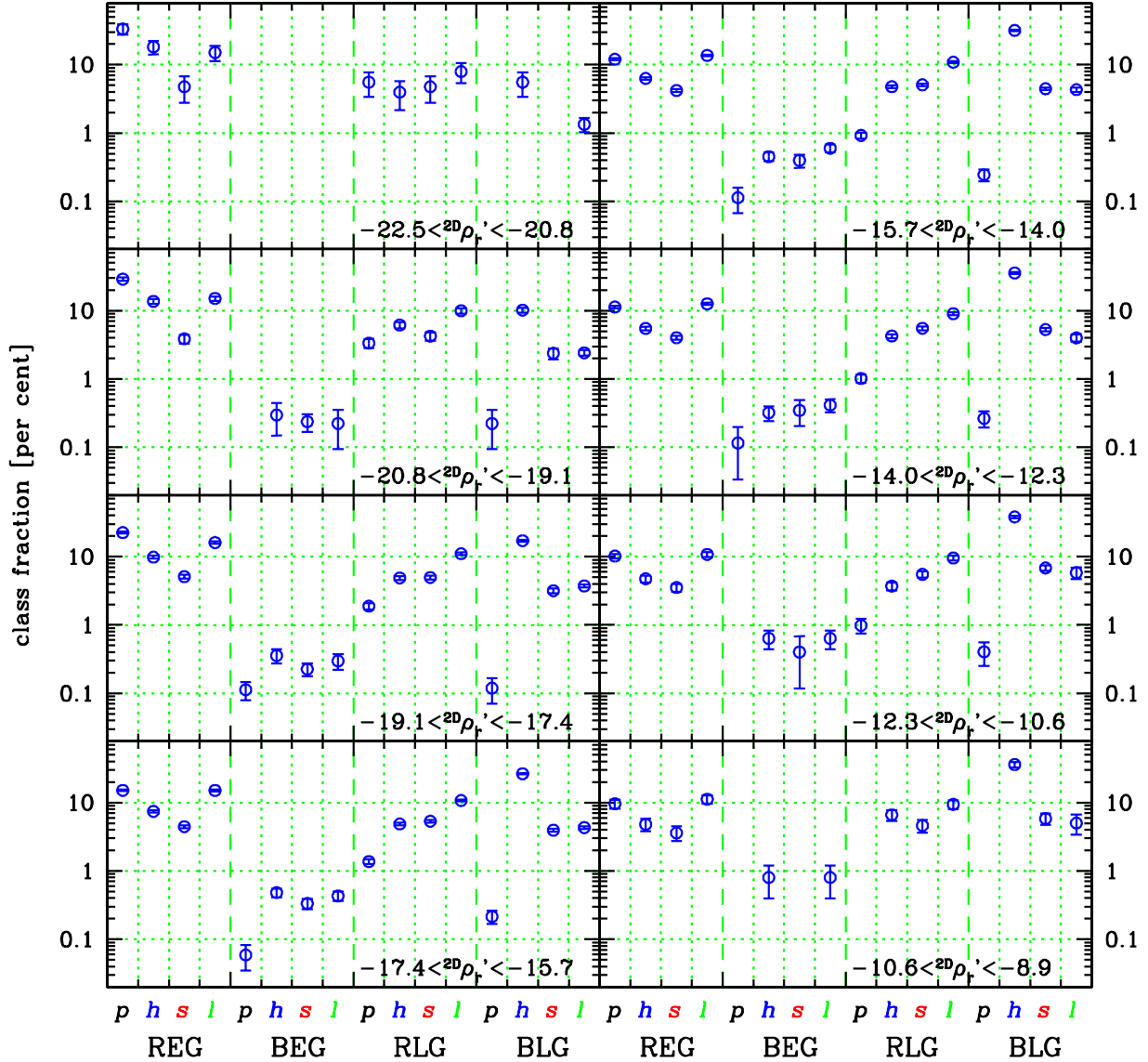}
\caption{ Comparison of the class fraction between the fine classes at given 2D local luminosity density bin. In each panel, the X-axis bins indicate the fine classes in order of: $p$REG, $h$REG, $s$REG, $l$REG, $p$BEG, $h$BEG, $s$BEG, $l$BEG, $p$RLG, $h$RLG, $s$RLG, $l$RLG, $p$BLG, $h$BLG, $s$BLG and $l$BLG. The errorbars show the Poisson errors. }
\label{clfracad}
\end{figure*}

\begin{figure}
\includegraphics[width=84mm]{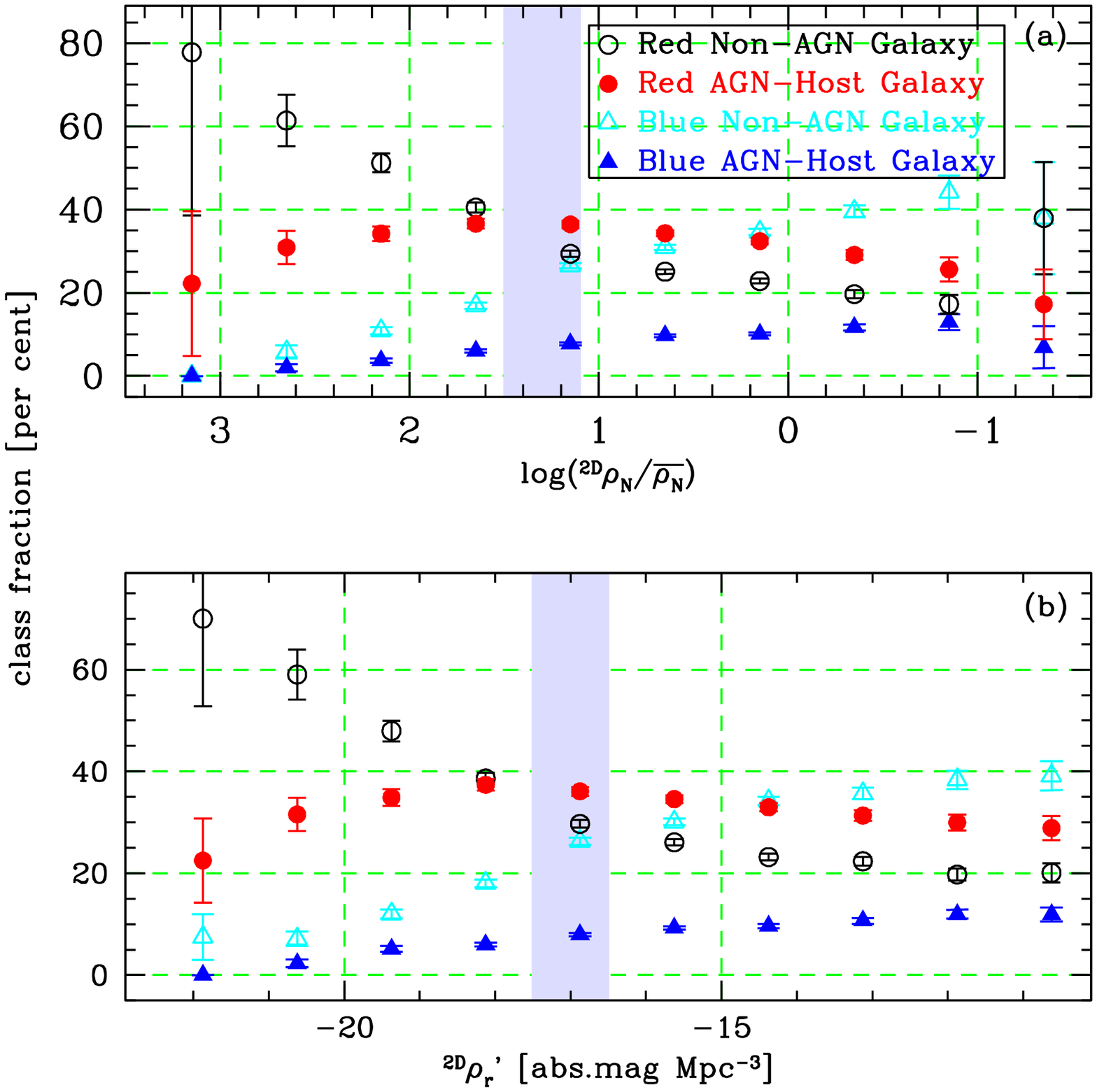}
\caption{ (a) The number fraction of red non-AGN galaxies (open circle), red AGN-host galaxies (filled circle), blue non-AGN galaxies (open triangle) and blue AGN-host galaxies (filled triangle) as a function of the 2D local number density. (b) The same as (a), but as a function of the 2D luminosity density. Errorbars show the Poisson errors. The intermediate ${\rho_r}'$ range between cluster and field environments is denoted as the shaded domain. }
\label{agnfrac}
\end{figure}

\section{Results}

\subsection{Local density}

Fig.~\ref{clfrac} and \ref{clfrac3d} show the dependence of each class fraction on the local number density and the (target-excluded) local luminosity density.
The class fraction in Fig.~\ref{clfrac} and \ref{clfrac3d} is defined as the number fraction of the galaxies in a fine class among the entire galaxies, at given local density bin.
The REG fraction at high-density environments is much larger than that at low-density environments. However, as reported by \citet{wes07}, it is found that AGN REGs ($s$REGs and $l$REGs) prefer slightly lower density environments than non-AGN REGs ($p$REGs and $h$REGs). In Fig.~\ref{clfrac}a, non-AGN REG fractions monotonically increase as the local density increases, while AGN REG fractions have a peak at $^{\textrm{\protect\tiny 2D}}{\rho_N}/\bar{\rho}_N\sim60$ ($^{\textrm{\protect\tiny 2D}}{\rho_r}'\sim-18$). At denser environments than the peak density, the number fraction of AGN REGs does not increase any more or even decreases with large uncertainty as the local density increases.
It is noted that the peak local density is very close to the intermediate local density between cluster and field environments (see Fig.~\ref{cluster}). This feature is hardly found in Fig.~\ref{clfrac3d}, but the 3D estimation is less reliable than the 2D estimation at high-density environments.
Note that the Y-axis scales in Fig.~\ref{clfrac} and Fig.~\ref{clfrac3d} are very different between the panels (i.e. between the fine classes). A better view of the fine class composition at fixed local density is presented in Fig.~\ref{clfracad}, which compares the class fraction between the fine classes at given 2D local luminosity density bin.

Non-AGN RLGs ($p$RLGs and $h$RLGs) show a trend similar to non-AGN REGs, in the sense that their number fraction increases as the local density increases. The fraction of AGN RLGs ($s$RLGs and $l$RLGs) shows a peak at intermediate-density environments and the number fraction of AGN RLGs decreases as the local density increases at high-density environments ($^{\textrm{\protect\tiny 2D}}{\rho_N}/\bar{\rho}_N>20$ or $^{\textrm{\protect\tiny 2D}}{\rho_r}'<-17$).
In the 3D local density, this feature is also found in the $l$RLGs but not found in the $s$RLGs (Fig.~\ref{clfrac3d}).
Fig.~\ref{agnfrac} compares the number fraction of red non-AGN galaxies ($p$REG + $h$REG + $p$RLG + $h$RLG), red AGN-host galaxies ($s$REG + $l$REG + $s$RLG + $l$RLG), blue non-AGN galaxies ($p$BEG + $h$BEG + $p$BLG + $h$BLG), and blue AGN-host galaxies ($s$BEG + $l$BEG + $s$BLG + $l$BLG).
In this figure, it is clearly shown that the number fraction of AGN-host galaxies rapidly decreases as the local density increases at high-density environments (which approximately correspond to galaxy cluster environments), with significant difference between red non-AGN galaxies and red AGN-host galaxies.

In Fig.~\ref{clfrac} and \ref{clfrac3d}, no significantly different trend is found between $p$REGs and $h$REGs, which implies that the local density is not directly responsible for the star formation in $h$REGs.
The fraction of AGN BEGs ($s$BEGs and $l$BEGs) increases as the local density decreases, whereas the non-AGN BEG fraction seem to have a peak at intermediate-density environments with large uncertainty.
The BLG fraction at low-density environments is much larger than that at high-density environments. The $p$BLG fraction as a function of local density is less biased to low local density than the other BLGs, but the uncertainty is large.

\begin{figure*}
\includegraphics[width=80mm]{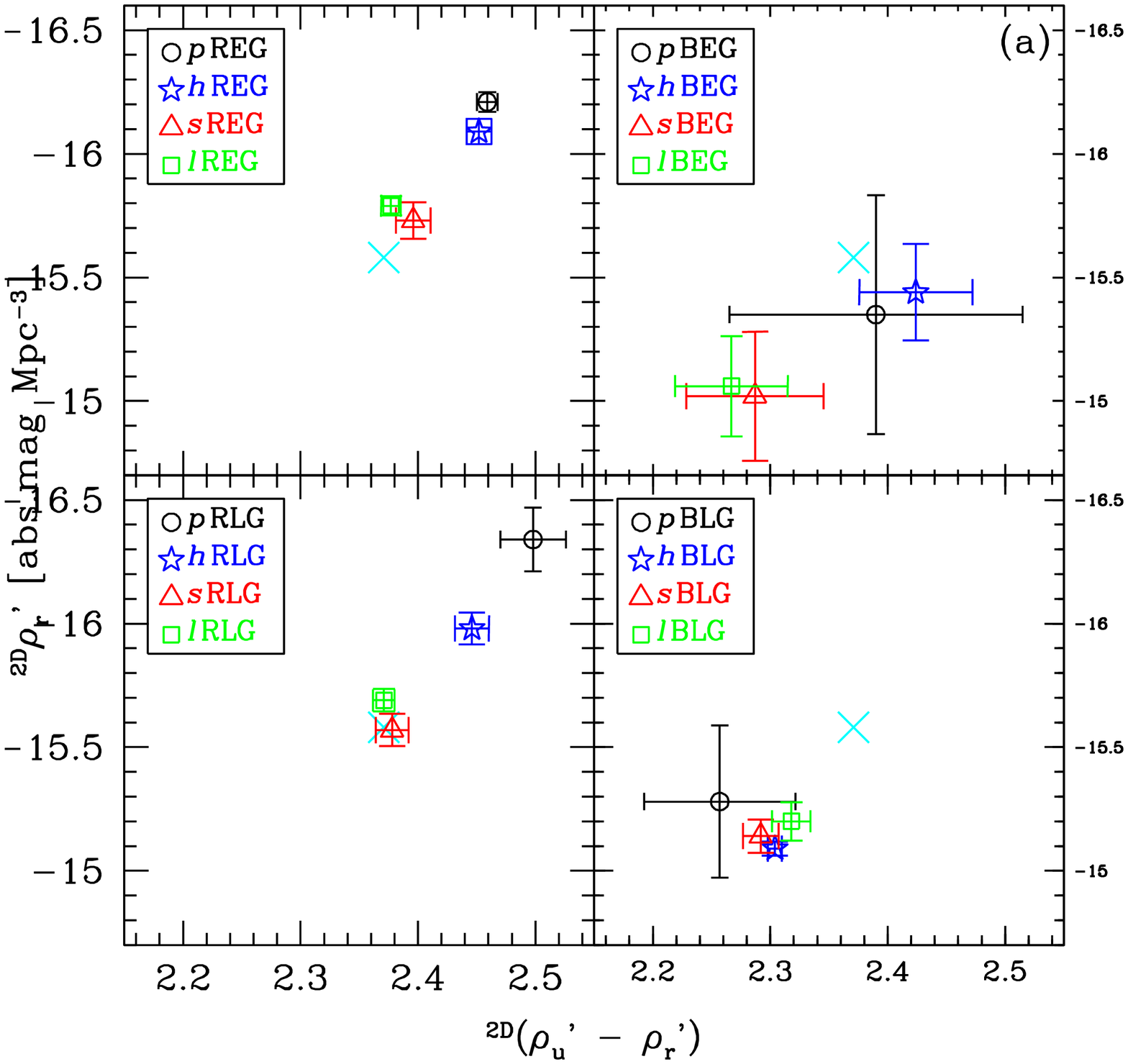}
\includegraphics[width=80mm]{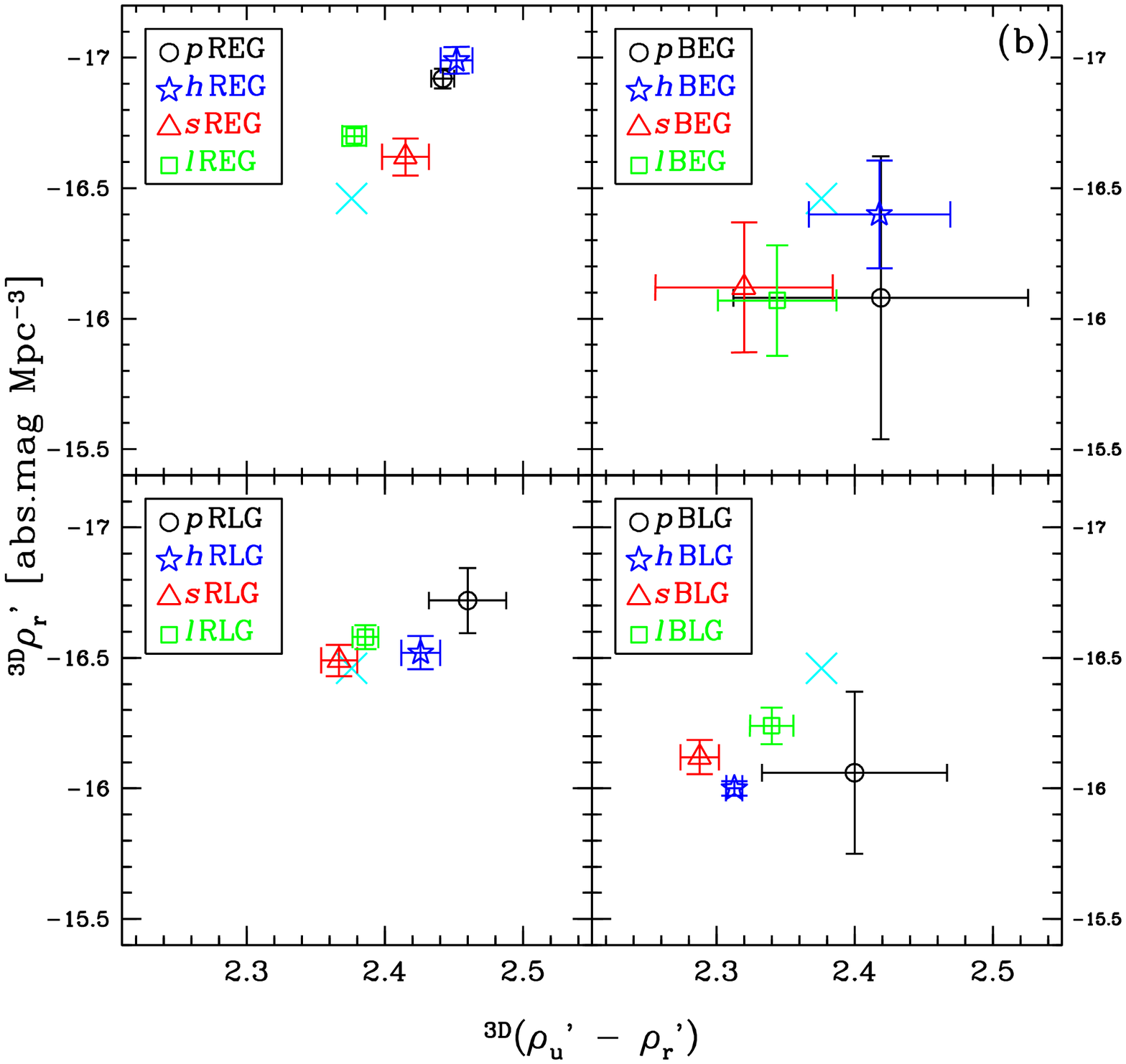}
\caption{ The target-excluded local luminosity density versus target-excluded local colour relation of galaxies divided into the 16 fine classes. Each symbol shows the median ${\rho_r}'$ and ${\rho_u}' - {\rho_r}'$ for each class and  and the errorbars indicate their sampling errors estimated from 200-times random sampling, using (a) 2D local density indicators and (b) 3D local density indicators. The cross symbol shows the median ${\rho_r}'$ and ${\rho_u}' - {\rho_r}'$ of the entire galaxies. }
\label{envcmr}
\end{figure*}

\begin{table}
\centering
\caption{Median 2D environmental parameters for the fine classes}
\label{envcmrtab}
\begin{tabular}{ccc}
\hline \hline
Class & $^{\textrm{\protect\tiny{2D}}}{\rho_r}'$ & $^{\textrm{\protect\tiny{2D}}}({\rho_u}' - {\rho_r}')$ \\
\hline
$p$REG & $-16.206\pm0.038$ (1.535) & $2.459\pm0.008$ (0.282) \\
$h$REG & $-16.089\pm0.056$ (1.463) & $2.452\pm0.011$ (0.277) \\
$s$REG & $-15.726\pm0.069$ (1.397) & $2.396\pm0.015$ (0.292) \\
$l$REG & $-15.786\pm0.035$ (1.374) & $2.377\pm0.009$ (0.274) \\
\hline
$p$BEG & $-15.347\pm0.507$ (1.330) & $2.390\pm0.111$ (0.215) \\
$h$BEG & $-15.443\pm0.217$ (1.229) & $2.424\pm0.044$ (0.245) \\
$s$BEG & $-15.020\pm0.259$ (1.463) & $2.287\pm0.056$ (0.253) \\
$l$BEG & $-15.062\pm0.206$ (1.152) & $2.267\pm0.044$ (0.296) \\
\hline
$p$RLG & $-16.335\pm0.128$ (1.678) & $2.498\pm0.026$ (0.300) \\
$h$RLG & $-15.982\pm0.067$ (1.425) & $2.446\pm0.014$ (0.275) \\
$s$RLG & $-15.567\pm0.058$ (1.413) & $2.378\pm0.014$ (0.269) \\
$l$RLG & $-15.692\pm0.043$ (1.355) & $2.371\pm0.009$ (0.276) \\
\hline
$p$BLG & $-15.277\pm0.282$ (1.575) & $2.257\pm0.062$ (0.335) \\
$h$BLG & $-15.093\pm0.027$ (1.291) & $2.304\pm0.006$ (0.292) \\
$s$BLG & $-15.142\pm0.068$ (1.407) & $2.292\pm0.016$ (0.297) \\
$l$BLG & $-15.196\pm0.074$ (1.257) & $2.318\pm0.017$ (0.281) \\
\hline \hline
\end{tabular}
\\
$\pm$values show the sampling errors, and (value)s show the SIQRs.
\end{table}

\begin{table}
\centering
\caption{Median 3D environmental parameters for the fine classes}
\label{envcmrtab2}
\begin{tabular}{ccc}
\hline \hline
Class & $^{\textrm{\protect\tiny{3D}}}{\rho_r}'$ & $^{\textrm{\protect\tiny{3D}}}({\rho_u}' - {\rho_r}')$ \\
\hline
$p$REG & $-16.923\pm0.035$ (1.349) & $2.442\pm0.008$ (0.279) \\
$h$REG & $-16.986\pm0.049$ (1.298) & $2.452\pm0.013$ (0.275) \\
$s$REG & $-16.624\pm0.065$ (1.289) & $2.415\pm0.016$ (0.291) \\
$l$REG & $-16.697\pm0.035$ (1.279) & $2.378\pm0.010$ (0.280) \\
\hline
$p$BEG & $-16.083\pm0.482$ (1.282) & $2.419\pm0.103$ (0.258) \\
$h$BEG & $-16.398\pm0.211$ (1.296) & $2.418\pm0.051$ (0.228) \\
$s$BEG & $-16.117\pm0.250$ (1.419) & $2.320\pm0.063$ (0.286) \\
$l$BEG & $-16.073\pm0.214$ (1.290) & $2.344\pm0.046$ (0.302) \\
\hline
$p$RLG & $-16.721\pm0.102$ (1.283) & $2.460\pm0.027$ (0.298) \\
$h$RLG & $-16.523\pm0.060$ (1.415) & $2.426\pm0.016$ (0.279) \\
$s$RLG & $-16.490\pm0.053$ (1.394) & $2.367\pm0.014$ (0.296) \\
$l$RLG & $-16.579\pm0.041$ (1.343) & $2.386\pm0.010$ (0.277) \\
\hline
$p$BLG & $-16.059\pm0.307$ (1.351) & $2.400\pm0.068$ (0.302) \\
$h$BLG & $-15.997\pm0.024$ (1.393) & $2.313\pm0.007$ (0.300) \\
$s$BLG & $-16.124\pm0.062$ (1.453) & $2.288\pm0.016$ (0.303) \\
$l$BLG & $-16.243\pm0.064$ (1.383) & $2.340\pm0.018$ (0.284) \\
\hline \hline
\end{tabular}
\\
$\pm$values show the sampling errors, and (value)s show the SIQRs.
\end{table}

\begin{figure*}
\includegraphics[width=80mm]{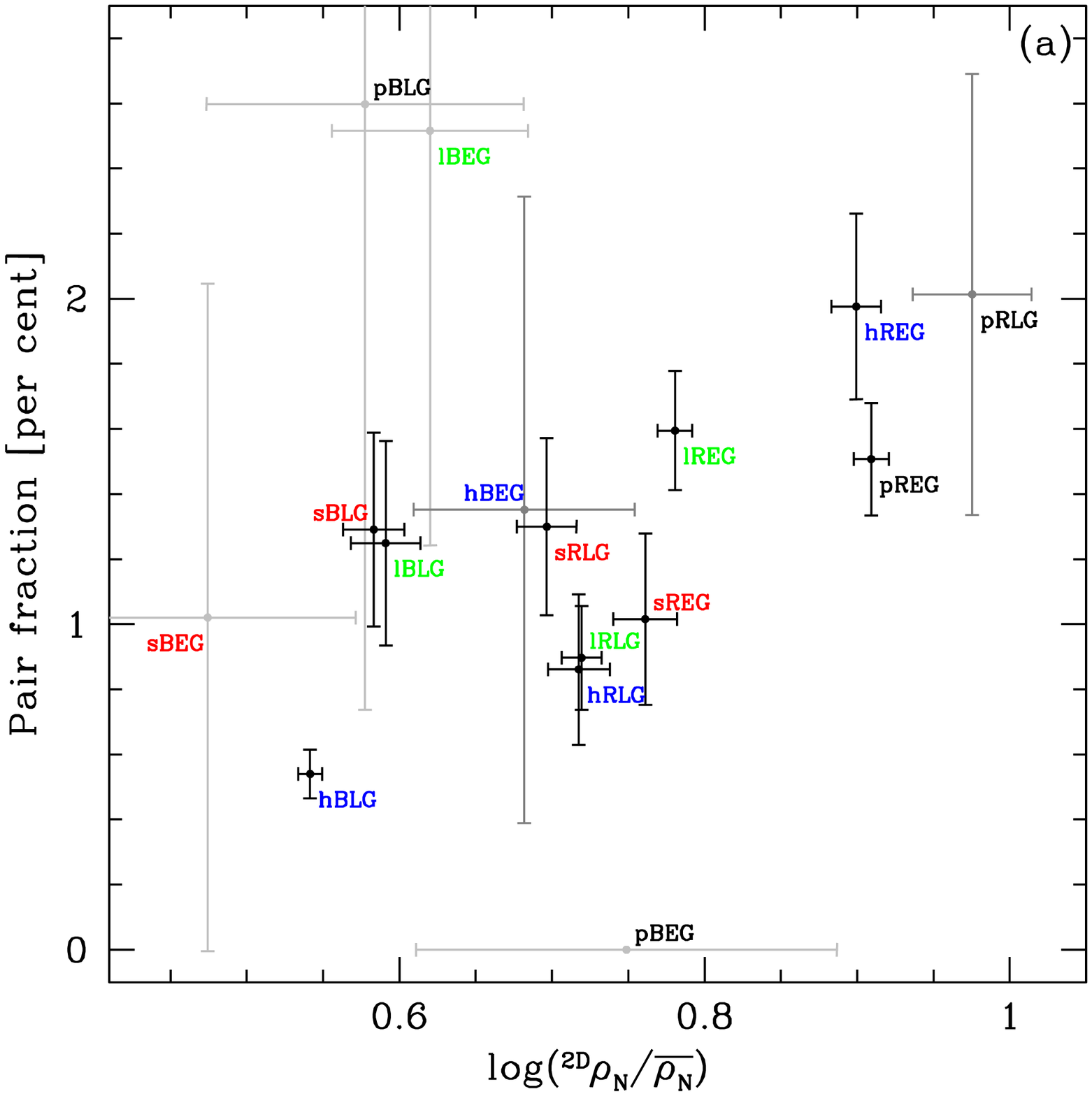}
\includegraphics[width=80mm]{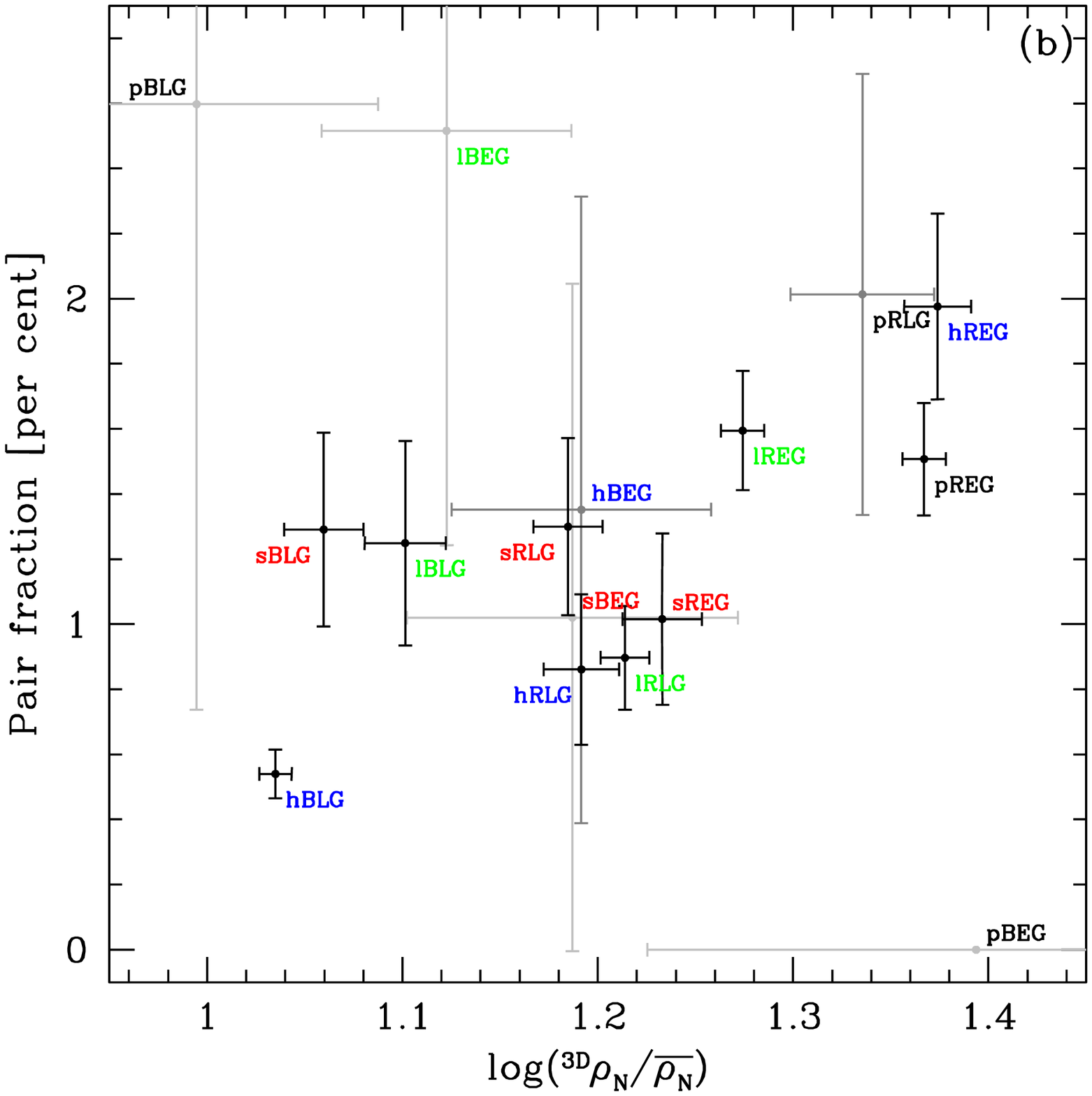}
\caption{ Close pair fraction versus (a) $^{\textrm{\protect\tiny 2D}}{\rho_N}/\bar{\rho}_N$ and (b) $^{\textrm{\protect\tiny 3D}}{\rho_N}/\bar{\rho}_N$. The errorbars show the Poisson errors of the close pair fraction and the sampling error of the ${\rho_N}/\bar{\rho}_N$. The datapoints with their Poisson errors $>1$ per cent are denoted using light grey, and those with $0.5\le$ Poisson error $<1$ per cent are denoted using dark grey. Zero per cent pair fraction is denoted using light grey. }
\label{pairfr}
\end{figure*}

\subsection{Median local luminosity density and the local colour}

Fig.~\ref{envcmr} plots the median ${\rho_r}'$ and ${\rho_u}' - {\rho_r}'$ of the 16 fine galaxy classes with their sampling errors. The sampling errors were estimated from 200-times random sampling tests for given sample size, which show how statistically significant the differences in the median ${\rho_r}'$ and ${\rho_u}' - {\rho_r}'$ are between different fine classes. The median ${\rho_r}'$ and ${\rho_u}' - {\rho_r}'$ values, their sample inter-quartile ranges (SIQRs) and sampling errors are summarised in Table \ref{envcmrtab} and \ref{envcmrtab2}.
In Fig.~\ref{envcmr}, even the fine classes in the same morphology-colour class have often significantly different environments. For example, $p$REGs and $h$REGs have similar environments ($^{\textrm{\protect\tiny{2D}}}{\rho_r}'\sim-16.2$ and $^{\textrm{\protect\tiny{2D}}}({\rho_u}'-{\rho_r}')\sim2.45$), but AGN REGs ($s$REGs and $l$REGs) have significantly different environments from those of non-AGN REGs by more than $5\sigma$, in the sense that AGN REGs prefer less dense and bluer environments ($^{\textrm{\protect\tiny{2D}}}{\rho_r}'\sim-15.7$ and $^{\textrm{\protect\tiny{2D}}}({\rho_u}'-{\rho_r}')\sim2.39$) than $p$REGs and $h$REGs.
AGN BEGs also prefer less dense and bluer environments than $h$BEGs with $2.3\sigma$ significance.

For late-type galaxies, the 2D and 3D environmental parameters show somewhat different trends from each other. For example, $p$RLGs prefer significantly higher local density (by $5.5\sigma$) and redder local colour (by $4.1\sigma$) than AGN RLGs in the 2D estimation, while the significance of such difference is relatively small in the 3D estimation ($2\sigma$ and $3.1\sigma$, respectively). The local density and the local colour of BLGs are almost the same between their different spectral classes in the 2D estimation. However, they show distinguishable difference in the 3D estimation, in the sense that $l$BLGs prefer slightly higher local density and redder local colour than $s$BLGs (by $2.5\sigma$).
As mentioned in \S\ref{C2d3d}, the 2D parameters show good performance at high-density environments, while the 3D parameters work well at low-density environments. Thus, it seems that the environments of RLGs are more reliable in the 2D estimation, whereas those of BLGs are more reliable in the 3D estimation.

$p$RLGs have environments similar to those of $p$REGs, but $p$BLGs prefer environments significantly ($3.3\sigma$) less dense than those of $p$REGs. The environments of $p$BEGs are intermediate between those of passive red galaxies and $p$BLGs, but their sampling errors are very large ($< 1\sigma$).
As mentioned in the previous section, $h$REGs prefer high-density environments in spite of their current star formation. $h$RLGs have the local density similar to those of $h$REGs, and the environments of $h$BEGs are less dense than those of $h$RLGs by $2.4\sigma$. It is noted that $h$REGs, $h$BEGs and $h$RLGs have different median ${\rho_r}'$ but similar median ${\rho_u}'-{\rho_r}'$. However, $h$BLGs prefer significantly ($>2.7\sigma$) bluer environments than the other H{\protect\scriptsize II} classes.
No large difference is found between Seyfert galaxies and LINER galaxies. Among AGN host galaxies, AGN REGs prefer the highest-density environments, and AGN RLGs have environments slightly ($1.8\sigma$) less dense than AGN REGs. AGN BEGs and AGN BLGs prefer similarly low-density environments.

\begin{figure*}
\includegraphics[width=80mm]{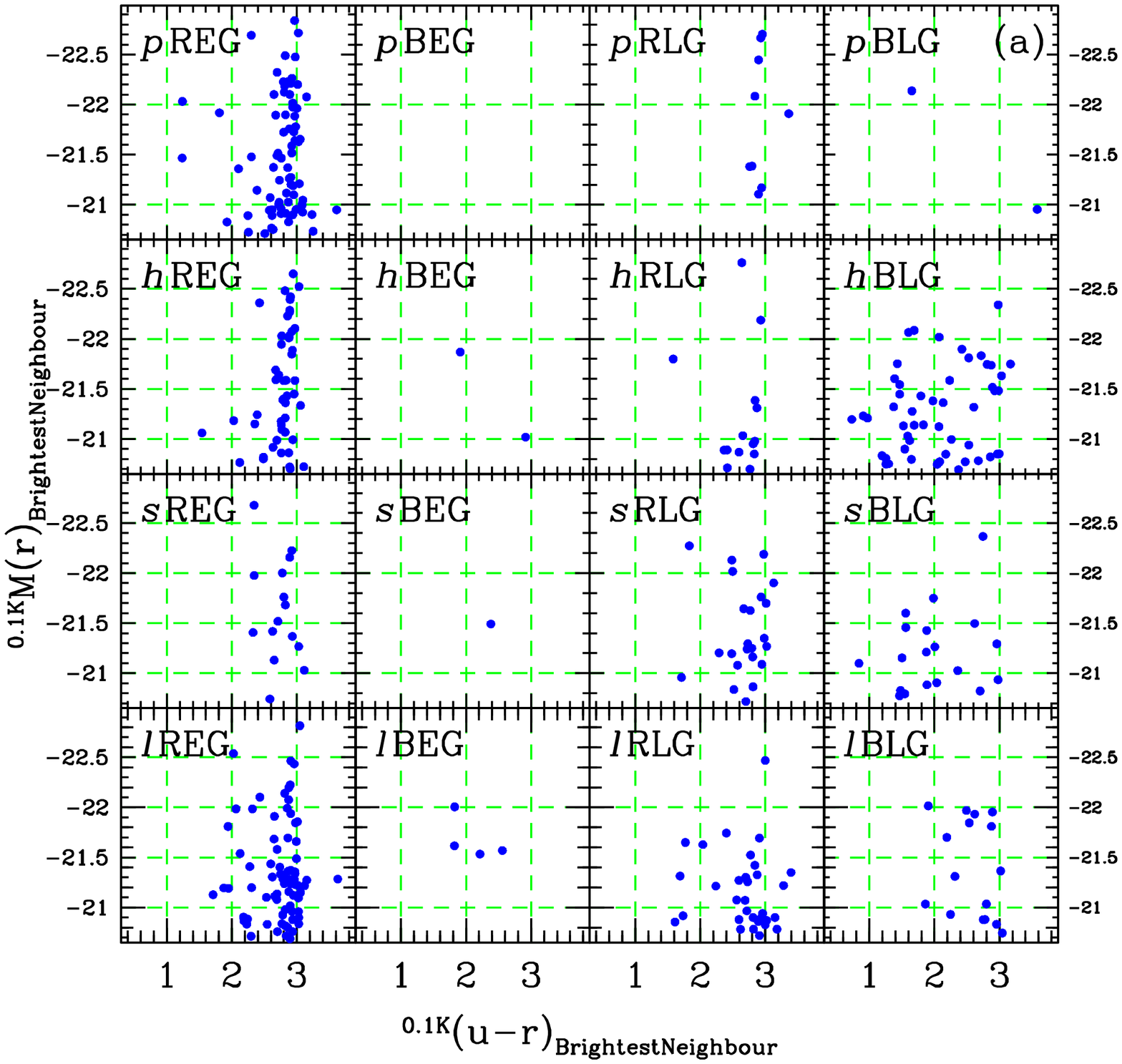}
\includegraphics[width=80mm]{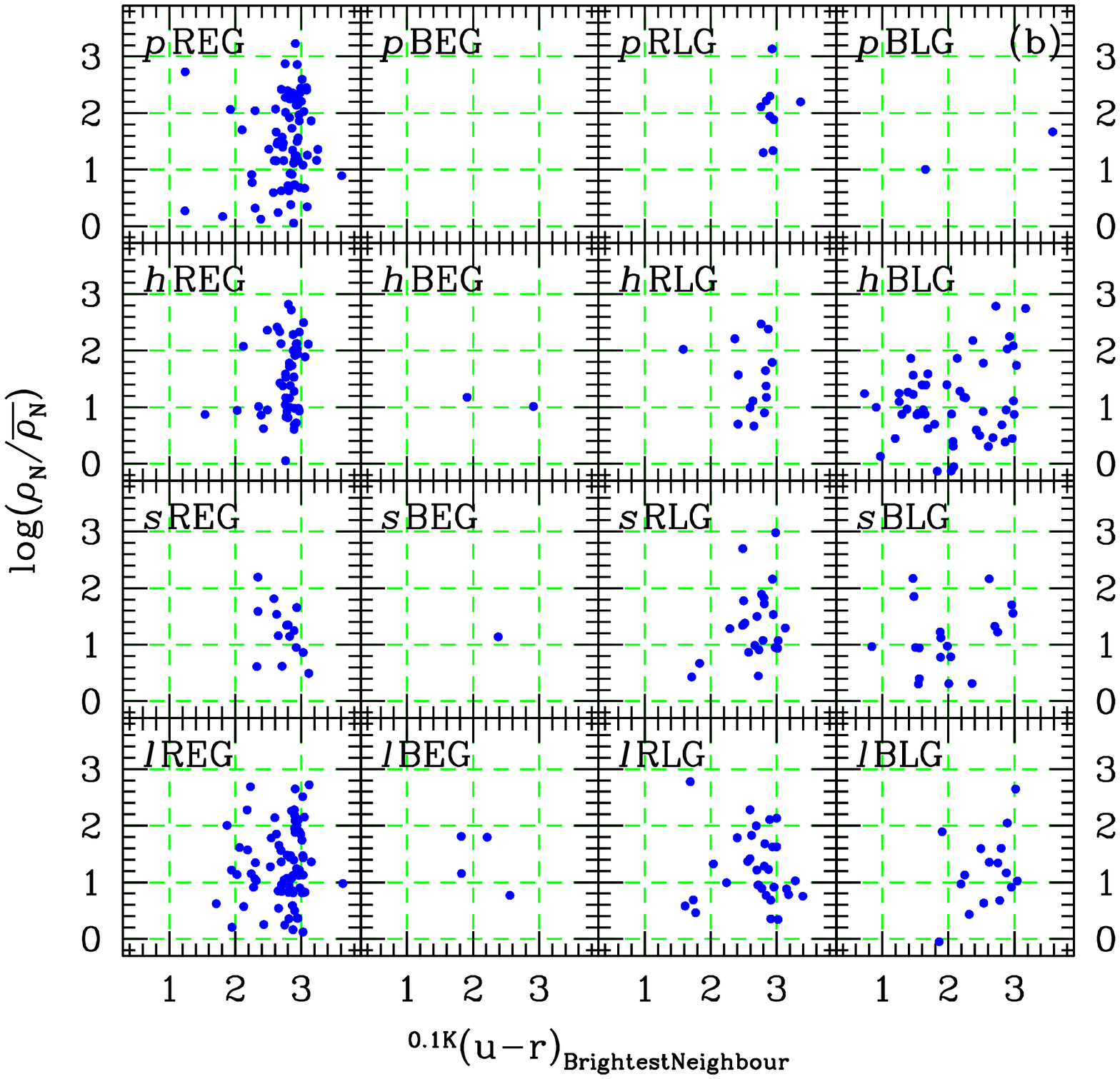}
\caption{ (a) The colour-magnitude relation and (b) the colour-density relation of the brightest neighbours in each fine class. }
\label{paircmr}
\end{figure*}

\begin{figure*}
\includegraphics[width=80mm]{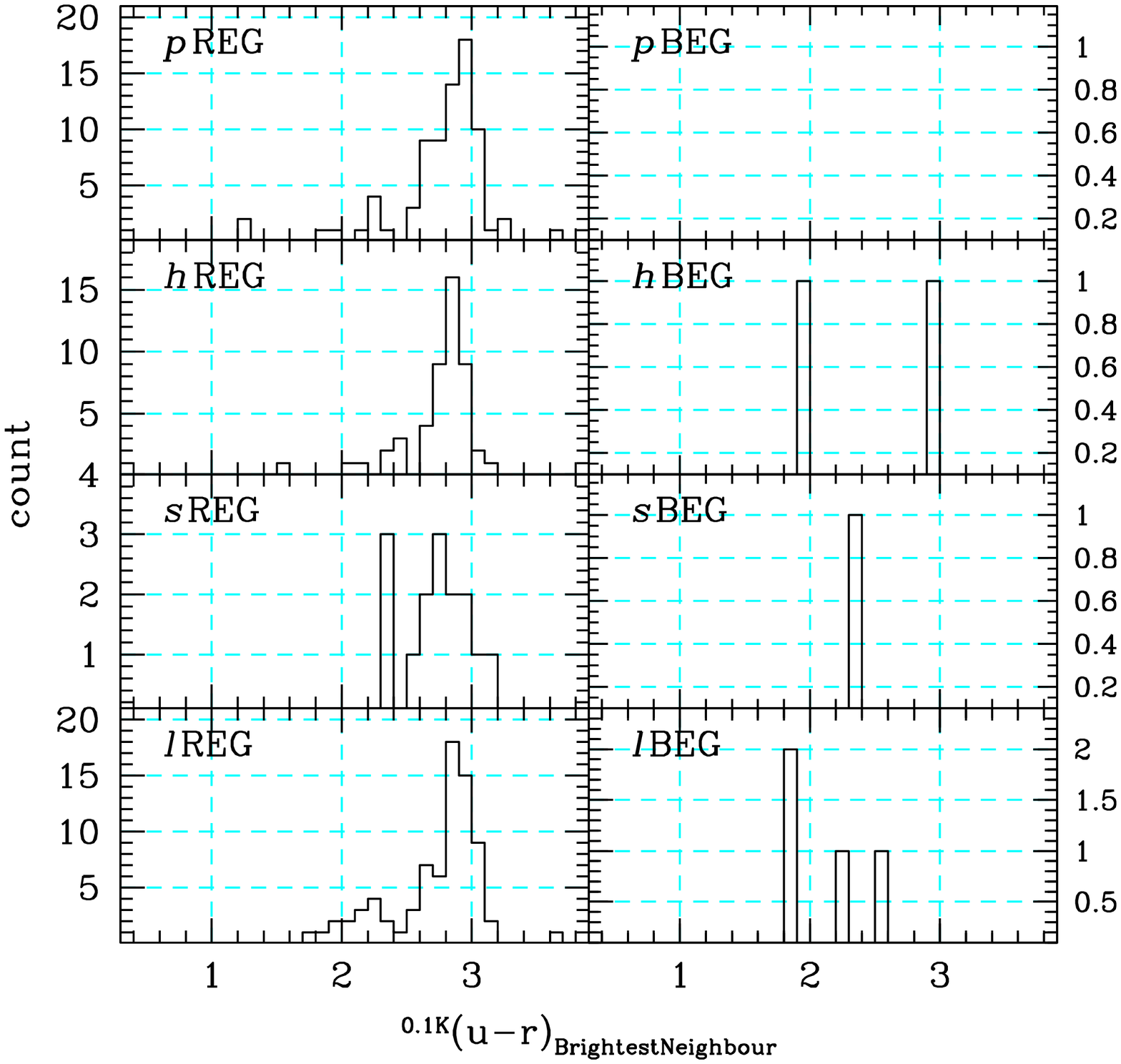}
\includegraphics[width=80mm]{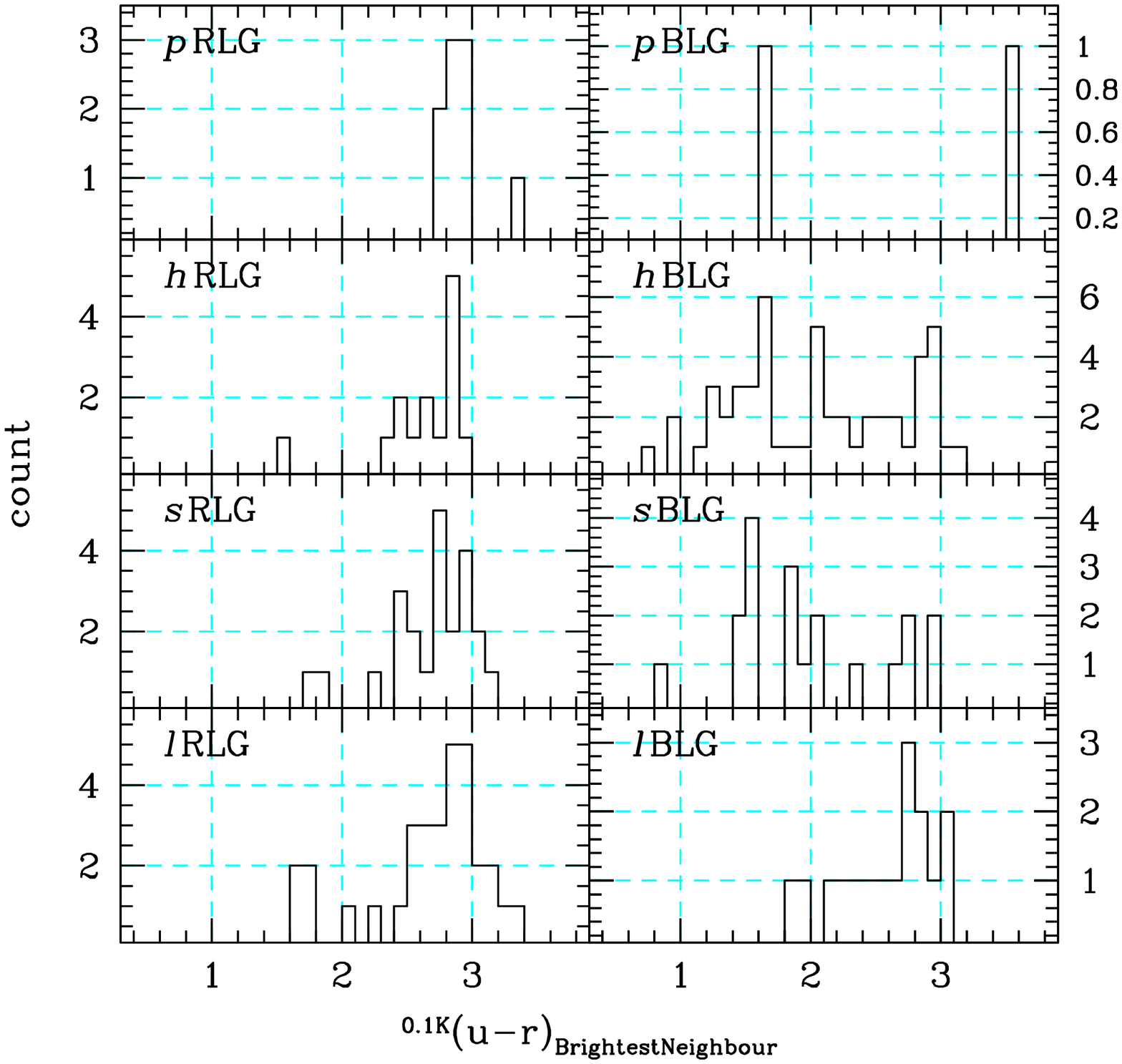}
\caption{ The colour distribution of the brightest neighbours in each fine class. }
\label{pairur}
\end{figure*}

\begin{table}
\centering
\caption{Close pair information for the fine classes}
\label{pairtab}
\begin{tabular}{ccc}
\hline \hline
Class & pair fraction$^{(a)}$ & $^{0.1{\textrm{\protect\scriptsize K}}}(u-r)_{\textrm{\protect\scriptsize BrightestNeighbour}}^{(b)}$ \\
\hline
$p$REG & $1.51\pm0.17\%$ & $2.86\pm0.15$ \\
$h$REG & $1.98\pm0.29\%$ & $2.82\pm0.09$ \\
$s$REG & $1.02\pm0.26\%$ & $2.78\pm0.17$ \\
$l$REG & $1.59\pm0.18\%$ & $2.85\pm0.17$ \\
\hline
$p$BEG & $0\%$ & --- \\
$h$BEG & $1.35\pm0.96\%$ & $2.92\pm0.50$ \\
$s$BEG & $1.02\pm1.03\%$ & $2.38\pm0.00$ \\
$l$BEG & $2.52\pm1.27\%$ & $2.21\pm0.37$ \\
\hline
$p$RLG & $2.01\pm0.68\%$ & $2.90\pm0.05$ \\
$h$RLG & $0.86\pm0.23\%$ & $2.76\pm0.21$ \\
$s$RLG & $1.30\pm0.27\%$ & $2.73\pm0.22$ \\
$l$RLG & $0.90\pm0.16\%$ & $2.81\pm0.18$ \\
\hline
$p$BLG & $2.60\pm1.86\%$ & $3.58\pm0.96$ \\
$h$BLG & $0.54\pm0.08\%$ & $2.07\pm0.57$ \\
$s$BLG & $1.29\pm0.30\%$ & $1.89\pm0.54$ \\
$l$BLG & $1.25\pm0.31\%$ & $2.75\pm0.29$ \\
\hline \hline
\end{tabular}
\\
\begin{itemize}
 \item[(a)] Close pair fraction. $\pm$value shows the Poisson error.
 \item[(b)] Median $^{0.1{\textrm{\protect\scriptsize K}}}(u-r)$ colour of the brightest neighbours. $\pm$value shows the SIQR.
\end{itemize}
\end{table}

\begin{figure*}
\includegraphics[width=80mm]{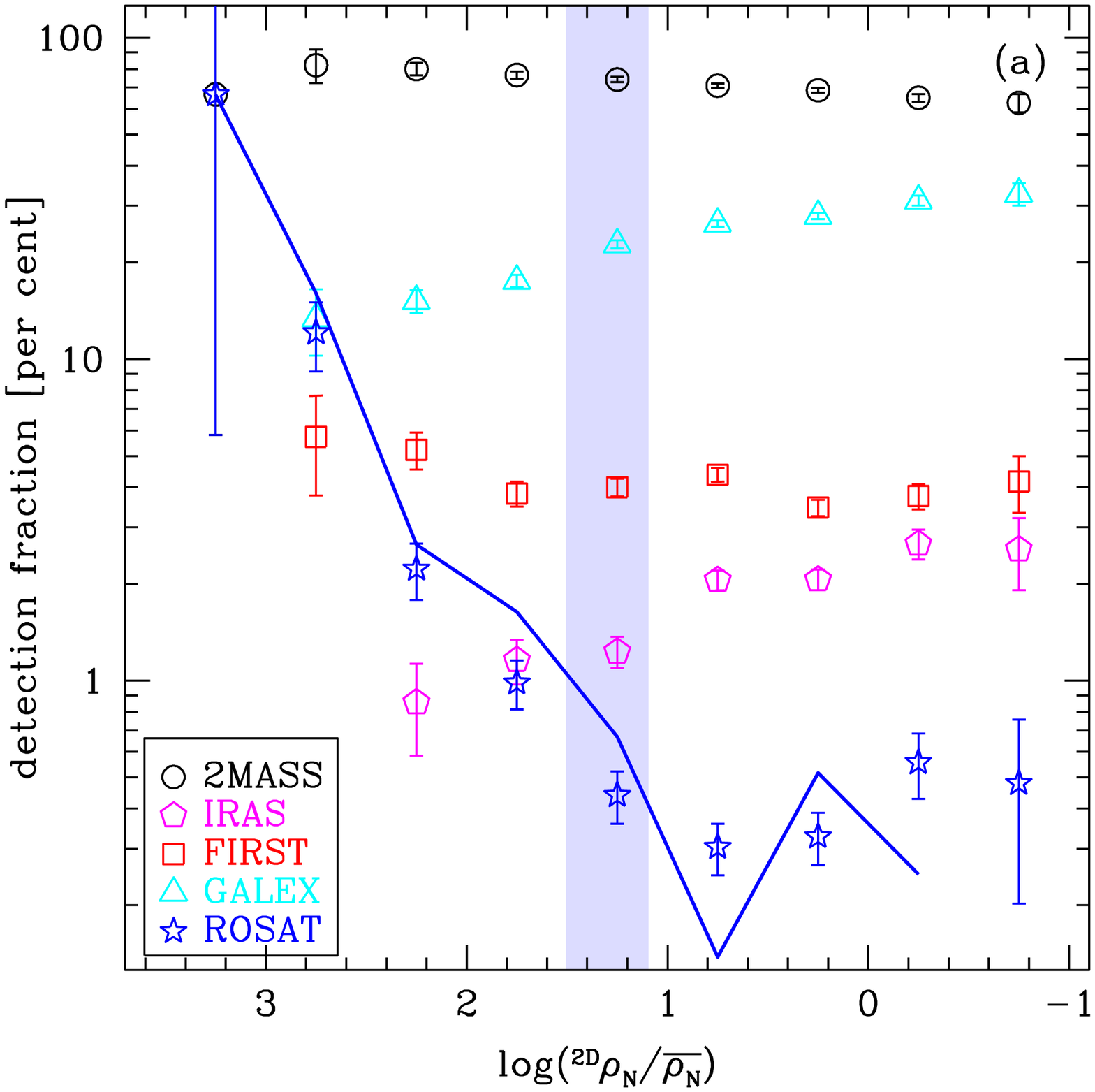}
\includegraphics[width=80mm]{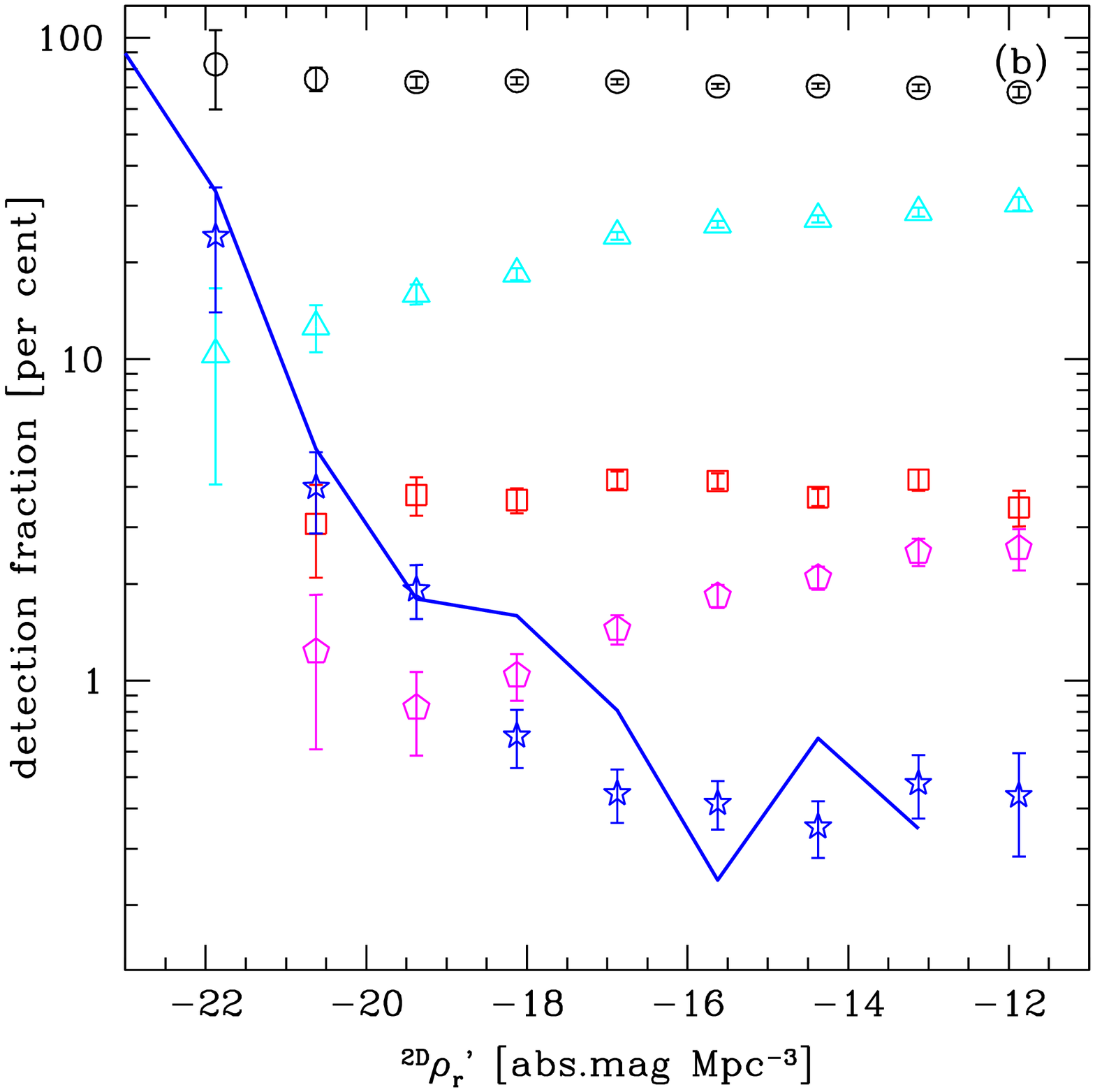}
\includegraphics[width=80mm]{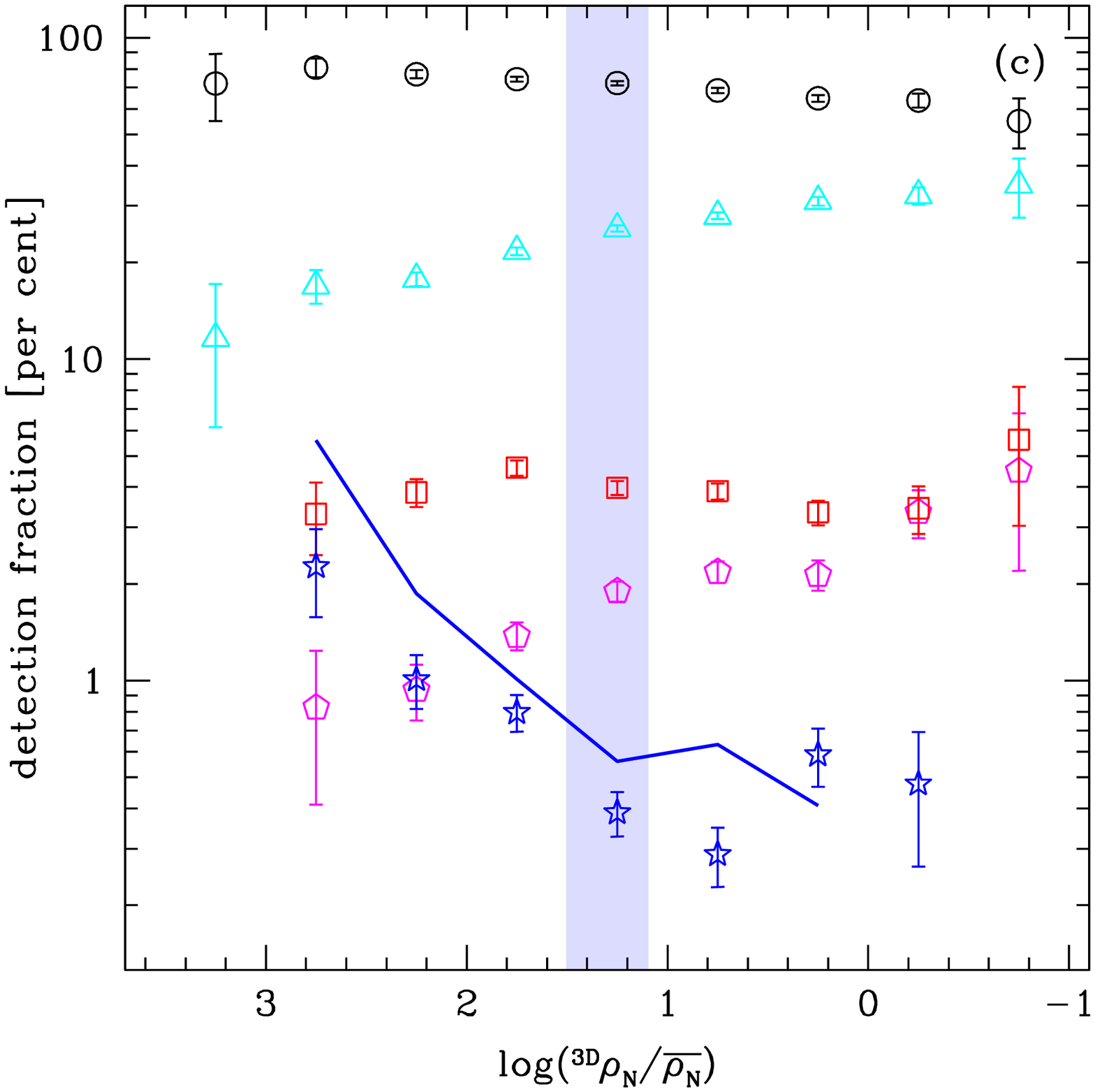}
\includegraphics[width=80mm]{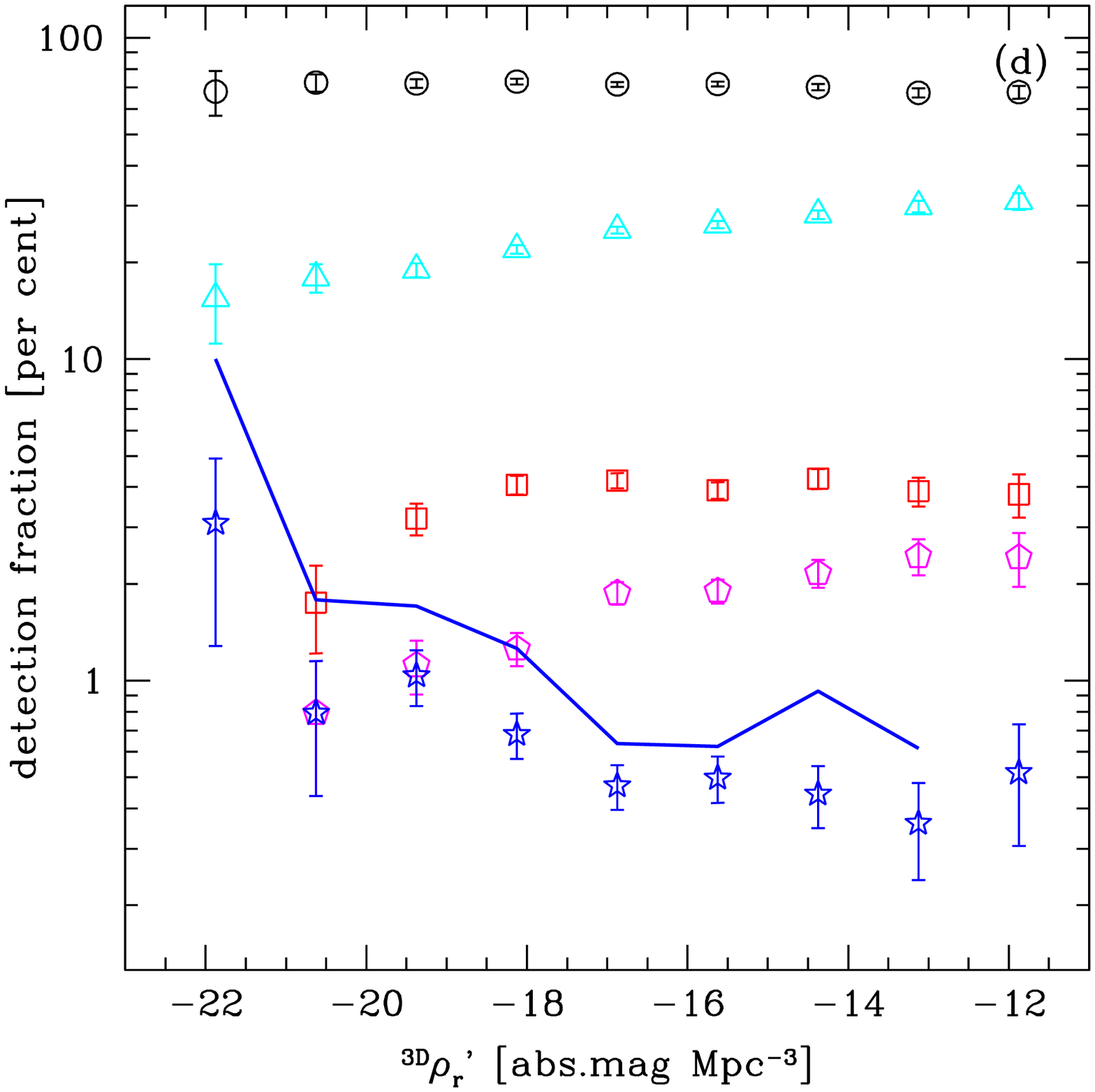}
\caption{ The dependence of the SDSS galaxy detection fractions in the 2MASS (circle), IRAS (pentagon), FIRST (rectangle), GALEX (triangle) and ROSAT (star) on (a) the 2D local number density, (b) the 2D local luminosity density, (c) the 3D local number density and (d) the 3D local luminosity density. The line shows the ROSAT detection fraction of the optically passive galaxies. The intermediate $\rho_N$ and ${\rho_r}'$ ranges between cluster and field environments are denoted as the shaded domains. The errorbars show the Poisson errors. }
\label{muldet}
\end{figure*}

\subsection{Close pair}

Fig.~\ref{pairfr} shows the relationship between the pair fraction and the local number density of the fine classes. As expected, it is found that the fine class with higher median local number density tends to have larger pair fraction in Fig.~\ref{pairfr}.
For example, the pair fraction of $h$REGs ($^{\textrm{\protect\tiny 2D}}{\rho_N}/\bar{\rho}_N\sim0.9$) is about 2 per cent, while that of $h$BLGs ($^{\textrm{\protect\tiny 2D}}{\rho_N}/\bar{\rho}_N\sim0.55$) is about 0.55 per cent.
In addition to this overall trend, some fine classes seem to show distinguishable difference in their pair fraction at given median local number density. For example, the pair fraction of $h$BLGs is half of that of AGN BLGs despite their similar local number density, which shows the possibility that the pair interaction is responsible for the AGN activity in BLGs.
Some other class sets seem to show pair fraction difference with similar local number density, such as `$s$RLGs -- $l$RLGs -- $h$RLGs', `$l$REGs -- $s$REGs' and `$h$REGs -- `$p$REGs', but their uncertainty is large.

Fig.~\ref{paircmr} presents what properties the brightest neighbours in each fine class have: absolute magnitude, colour and local number density. It is found that the brightest neighbours of red galaxies have a narrow range of their colour, while those of blue galaxies show scattered colour distributions. This difference does not seem to correlate with the absolute magnitude or the local number density.
Fig.~\ref{pairur} shows the $^{0.1{\textrm{\protect\scriptsize K}}}(u-r)$ colour distribution of the brightest neighbours in each fine class.
Red galaxies show similar $^{0.1{\textrm{\protect\scriptsize K}}}(u-r)_{\textrm{\protect\scriptsize BrightestNeighbour}}$ distributions regardless of their fine class: a single peak at $^{0.1{\textrm{\protect\scriptsize K}}}(u-r)_{\textrm{\protect\scriptsize BrightestNeighbour}}=2.8$ and a small blue tail. On the other hand, $h$BLGs and $s$BLGs show quite different $^{0.1{\textrm{\protect\scriptsize K}}}(u-r)_{\textrm{\protect\scriptsize BrightestNeighbour}}$ distributions from those of the red galaxies: the $^{0.1{\textrm{\protect\scriptsize K}}}(u-r)_{\textrm{\protect\scriptsize BrightestNeighbour}}$ distributions of $h$BLGs and $s$BLGs show double peak features, with the primary blue peak at $^{0.1{\textrm{\protect\scriptsize K}}}(u-r)_{\textrm{\protect\scriptsize BrightestNeighbour}}=1.6$ and the secondary red peak at $^{0.1{\textrm{\protect\scriptsize K}}}(u-r)_{\textrm{\protect\scriptsize BrightestNeighbour}}=2.8$. The secondary red peak colour of the $h$BLG and $s$BLG brightest neighbours is similar to the single peak colour of the red galaxy brightest neighbours.
It is noted that the blue peak of $s$BLGs is twice in size of their red peak, whereas $l$BLGs show a single red peak.
Table \ref{pairtab} lists the pair fraction and median $^{0.1{\textrm{\protect\scriptsize K}}}(u-r)$ colour of the brightest neighbours in each class.

\subsection{Multi-wavelength detection and environments}

In Paper II, we presented several multi-wavelength properties of the SDSS galaxies divided into fine classes. In the multi-wavelength study, the detection trends of the fine classes vary largely according to the survey wavelength.
For example, the most/least efficiently detected classes are $l$RLGs/$p$BEGs in the 2MASS, while those in the FIRST are $s$BLGs/$p$RLGs. Such variation is mainly due to the differences in the internal properties between the fine classes, but some galaxies show multi-wavelength properties that are not easy to understand just by their intrinsic characteristics, such as optically passive but X-ray bright galaxies.
Those unusual properties are sometimes related to their external conditions, because some environments (e.g. galaxy clusters) themselves have strong emission in some wavelength such as X-ray \citep{bou04}.
Thus, in this section, we briefly investigate the environmental dependence of multi-wavelength detection fractions for our sample galaxies, which helps us to understand the multi-wavelength behaviours of the fine classes better.

Fig.~\ref{muldet} shows the dependence of the detection fractions in the multi-wavelength surveys on the local luminosity density.
The 2MASS detection fraction increases as the local density increases, whereas the detection fractions in the IRAS and GALEX decrease with increasing local density. Considering the detection trends of the fine galaxy classes in the multi-wavelength bands (Paper II), Fig.~\ref{muldet} confirms the previous finding that galaxies in high-density environments tend to have red colours, old ages and early-type morphology, compared to those in low-density environments \citep{bla05,ber06,par07}.
The FIRST detection fraction shows quite different trends between the 2D and 3D local density parameters. The 2D local density of the FIRST detection fraction increases almost monotonically as the local density increases, while its 3D local density shows a peak at $^{\textrm{\protect\tiny 3D}}{\rho_N}/\bar{\rho}_N\sim60$ ($^{\textrm{\protect\tiny 3D}}{\rho_r}'\sim-18$). Considering that the 2D local density is more reliable than the 3D local density at high-density environments, it seems like that the galaxies at higher-density environments have higher probability to be radio sources.

The ROSAT detection fraction shows a very strong dependence on local luminosity density. In Fig.~\ref{muldet}a, the largest (at $^{\textrm{\protect\tiny 2D}}{\rho_N}/\bar{\rho}_N\sim1700$) and the smallest (at $^{\textrm{\protect\tiny 2D}}{\rho_N}/\bar{\rho}_N\sim5$) ROSAT detection fractions are different by 2.3 dex.
This result indicates that the bias of the ROSAT-detected galaxies to red colour (Paper II) is closely related to the environments of galaxies, because red galaxies prefer high-density environments as shown in Fig.~\ref{clfrac}. In Fig.~\ref{muldet}, the ROSAT detection fraction of the \emph{passive} galaxies (solid line) is very similar to that of the entire galaxies (open circles), showing that the passive galaxies in high-density environments have high probability to be detected in the X-ray.
Since galaxy clusters are typically strong X-ray emitters with a large amount of intracluster hot gas \citep{mcm89,rhe91,bou04}, some ROSAT-detected passive galaxies in high-density environments may be affected by such cluster environments.
However, some passive galaxies in low-density environments are also detected in the X-ray. The number fraction of the ROSAT-detected passive galaxies at $^{\textrm{\protect\tiny 3D}}{\rho_r}'>-17$ among the entire ROSAT-detected passive galaxies is about 30 per cent (16 of 54), which are classified as X-ray Bright Optically Normal Galaxies \citep[XBONGs;][]{yua04,civ07}.

\section{Discussion}

\subsection{Red Early-type Galaxies}\label{disreg}

The environments of REGs are previously well studied, in the sense that galaxies at high-density environments tend to have early-type morphology \citep{oem74,dre80,pos84}. Recently, \citet{par07} showed that the number fraction of early-type galaxies monotonically increases as the local density increases at given luminosity.
In this paper, $p$REGs and $h$REGs show good agreement with those previous results, displaying strong and monotonic dependence of their number fraction on their local density.
However, unlike those non-AGN REGs, the number fraction of AGN REGs ($s$REGs and $l$REGs) is not a monotonic function of the local density. Instead, they have a peak at $^{\textrm{\protect\tiny 2D}}{\rho_N}/\bar{\rho}_N\sim60$ ($^{\textrm{\protect\tiny 2D}}{\rho_r}'\sim-18$), which is very close to the intermediate local density between cluster and field environments shown in Fig.~\ref{cluster}.
This difference between non-AGN and AGN REGs results in the difference in their median local luminosity density and local colour (Fig.~\ref{envcmr}): AGN REGs prefer less dense and bluer environments than $p$REGs and $h$REGs by more than $5\sigma$ significance.
This result is closely related to the fact that AGN host galaxies prefer the \emph{green valley} domain in the colour-magnitude diagram, which is intermediate between the red and blue sequences \citep{sil08,cho09b,hic09}, because galaxy colour and local density are tightly correlated \citep{par07}.

If we suppose that galaxies at very-high-density environments are mostly cluster galaxies (supported by Fig.~\ref{cluster}), cluster environments seem to suppress AGN activity, reducing the number fraction of AGN REGs at very-high-density environments. \citet{par09b} discussed several possible mechanisms causing a sudden variation of the galaxy morphology in cluster environments, such as galaxy-galaxy tidal interaction, galaxy-cluster potential interaction and galaxy-galaxy hydrodynamic interaction.
Those mechanisms probably affect also the AGN activity in cluster environments: the mechanisms such as galaxy-galaxy interactions or harassment may deprive cluster galaxies of cold gas that can be the fuel of AGNs, resulting in a sudden decrease of AGN host galaxies in very-high-density environments.

During the study of the early-type star formation history, \citet{rog07} found that a significant number of galaxies in high-density environments undergo low but detectable recent star formation. $h$REGs in our sample may correspond to those objects, which show no significant difference in the local density distribution from $p$REGs. Such similarity of environments between $p$REGs and $h$REGs implies that the local density is not directly responsible for the star formation in $h$REGs.
Since satellite or close pair galaxies play an important role in determining the properties of the target galaxy \citep{van08,par08,par09a}, it is a possible scenario that the star formation in $h$REGs is triggered by the interaction with their close neighbour galaxies.
In Fig.~\ref{pairfr}, it is found that $h$REGs tend to have higher chance of pairs ($1.98\pm0.29$ per cent) than $p$REGs ($1.51\pm0.17$ per cent). However, this difference is statistically not meaningful ($1.4\sigma$ significance) in our sample.
It is an open question why the AGN-suppressing mechanisms (discussed in the previous paragraph) seem not to suppress the star formation in $h$REGs.

\subsection{Blue Early-type Galaxies}

Since the number fraction of BEGs out of the entire galaxies is very small (about 1.3 per cent in our sample), there are not many previous statistical studies on their environments. In \citet{par07}, it has been shown that the number fraction of BEGs out of early-type galaxies decreases as the local density increases. More recently, \citet{kan09} found that many blue-sequence E/S0 galaxies reside in low-to-intermediate-density environments.
Similar results are shown in this paper that the BEG fraction at low-density environments is higher than that at high-density environments.
However, there seems to be a difference between non-AGN BEGs and AGN BEGs, in the sense that AGN BEG ($s$BEG and $l$BEG) fraction increases monotonically as the local density decreases, whereas the non-AGN BEG fraction has a intermediate-density peak (Fig.~\ref{clfrac} and \ref{clfrac3d}).
This is similar to the relationship between non-AGN REGs and AGN REGs. However, due to the very small sample size of BEGs, the uncertainty is so large that the statistical reliability of the results for BEGs is not high (less than $2.3\sigma$ significance from Fig.~\ref{envcmr}).

In Paper I and Paper II, BEGs show the photometric and structural properties implying that they are early-type galaxies in the forming phase.
Since it is known that high-density environments accelerate galaxy evolution \citep{mat07}, the preference for low-density environments of BEGs seem to be natural, explaining why those early-type galaxies (BEGs) are forming so lately compared to REGs.
In detail, however, the environmental discrepancy between non-AGN BEGs and AGN BEGs is not easy to explain, because AGN-host galaxies are thought to be in the intermediate phase from star-forming galaxies to passive galaxies \citep[AGN feedback;][]{ant08,raf08}.

\subsection{Red Late-type Galaxies}

RLGs are mostly regarded as bulge-dominated late-type galaxies (Paper I and II). RLGs are not frequently mentioned in the previous studies, but several studies provided some hints on their environments. That is, it is known that galaxies have greater bulge-to-disc ratio (or high light-concentration) with increasing local density \citep{giu95,has99}.
As probable bulge-dominated galaxies, RLGs agree with those previous findings.
Non-AGN RLGs ($p$RLGs and $h$RLGs) show a trend similar to non-AGN REGs, in the sense that their number fraction increases as the local density increases.
On the other hand, the AGN RLG ($s$RLG and $l$RLG) fraction shows a peak at intermediate-density environments, which is a trend similar to that of AGN REGs.
In the same context as AGN REGs (\S\ref{disreg}), it seems that the AGN activity in RLGs are suppressed at very-high-density environments.

The unanswered problem in \S\ref{disreg} (why very-high-density environments do not suppress the star formation?) is less serious in RLGs, because the number fraction of $h$RLGs slightly decreases at the highest-density environment with large uncertainty (Fig.~\ref{clfrac}; $1.2\sigma$ significance). However, this decrease is statistically not meaningful and the star formation in $h$RLGs seems to be clearly less suppressed than the AGN activity in AGN RLGs at high-density environments. To understand this problem better, further studies on the H{\protect\scriptsize II} galaxies and AGN host galaxies in cluster environments are needed in the future.

In Paper I and II, $p$RLGs have properties very similar to those of REGs. Interestingly, in this paper, $p$RLGs have environments similar to those of $p$REGs. This result supports the idea that $p$RLGs are intermediate objects between early-type galaxies and late-type galaxies as suggested in Paper I and II.

\subsection{Blue Late-type Galaxies}

BLGs mostly consist of spiral or irregular galaxies, which have the opposite trend to REGs in their environments. That is, galaxies with late-type morphology tend to be at low-density environments \citep{oem74,dre80,pos84,par07}.
In this paper, consistent with those previous findings, the number fraction of BLGs increases almost monotonically as the local density decreases. Exceptionally, $p$BLGs seem to prefer intermediate density environments, but the sample size is too small.

In Fig.~\ref{pairfr}, $h$REGs have a similar pair fraction to $p$REGs. On the contrary, $h$BLGs have a smaller pair fraction than the other BLGs (significant up to $2.4\sigma$) despite their similar local number density. This result indicates that the close pair interaction is less important for the star formation of BLGs compared to that of REGs, at least in our sample luminosity range ($-22.92 < ^{0.1\textrm{\protect\scriptsize KE}}M_{pet}(r) < -20.71$). On the other hand, the close pair interaction may be responsible for the AGN activity in BLGs.

In Fig.~\ref{pairur}, the different spectral classes of BLGs show impressive differences in their close pair colour distribution.
$h$BLGs have both a blue peak ($\sim1.6$) and a red peak ($\sim2.8$) in their $^{0.1{\textrm{\protect\scriptsize K}}}(u-r)_{\textrm{\protect\scriptsize BrightestNeighbour}}$ distribution, and the two peaks have similar size.
$s$BLGs also show the double peak feature, but the red peak is just half in size compared to the blue peak.
On the other hand, $l$BLGs have only a red peak, which is similar to that of red galaxies.
The difference in the brightest neighbour colour between $s$BLGs and $l$BLGs also affect the median local colours of them, in the sense that $s$BLGs prefer slightly bluer local colour than $l$BLGs (Fig.~\ref{envcmr}b).
It is noted that the single red $^{0.1{\textrm{\protect\scriptsize K}}}(u-r)_{\textrm{\protect\scriptsize BrightestNeighbour}}$ peak of $l$BLGs is unusual, considering the conformity of pair galaxies \citep{wei06,par08}. These results show a possibility that the supply of fuel gas from outer sources is closely related to the AGN type: Seyfert or LINER. In other words, at least for BLGs, it seems like that Seyfert galaxies tend to have blue close pairs that may supply rich fuel gas for the target galaxy, unlike LINER galaxies.

However, this conclusion is very cautious for two reasons.
First, the uncertainty in the $^{0.1{\textrm{\protect\scriptsize K}}}(u-r)_{\textrm{\protect\scriptsize BrightestNeighbour}}$ distribution of BLGs is large, because the number of BLGs hosting close pair galaxies is not large enough.
Second, such a trend (the difference in the $^{0.1{\textrm{\protect\scriptsize K}}}(u-r)_{\textrm{\protect\scriptsize BrightestNeighbour}}$ distribution between Seyferts and LINERs) is not found in the other morphology-colour classes. This may be due to the limited range of the sample galaxy luminosity, but it is not assured in this paper.
Nevertheless, it is not easy to neglect the difference between $s$BLGs and $l$BLGs shown in Fig.~\ref{paircmr} and \ref{pairur}, either. This problem needs more inspection in the future, using a larger sample of close pairs.

\subsection{Comparison in a given spectral class}

$p$RLGs have environments similar to those of $p$REGs, but $p$BLGs prefer environments significantly less dense and bluer than those of $p$REGs ($3.3\sigma$ significance). The local environments of $p$BEGs are intermediate between those of passive red galaxies and $p$BLGs, but their sampling errors are large (less than $1\sigma$ significance).
The preference for low-density environments of $p$BLGs is not surprising, because many of them may not be genuinely passive (Paper II). Except for $p$BLGs, the other passive galaxies prefer relatively dense and red environments as expected.
All ROSAT-detected optically-passive galaxies are passive red galaxies (Paper II) and about 70 per cent of them reside in high-density environments like galaxy clusters. Since a galaxy cluster itself is a strong source of X-ray, some of those galaxies may be contaminated by the galaxy cluster X-ray emission.

In spite of their current star formation, $h$REGs prefer high-density environments. It is interesting that $h$REGs, $h$BEGs and $h$RLGs have different ${\rho_r}'$ distributions but similar ${\rho_u}'-{\rho_r}'$ distributions. On the other hand, $h$BLGs prefer significantly (more than $2.7\sigma$ significance) bluer environments than the other H{\protect\scriptsize II} classes.
Since the luminosity of our sample galaxies is limited to the bright end, this does not necessarily mean that H{\protect\scriptsize II} galaxies except $h$BLGs do not have blue neighbours. If $h$REGs, $h$BEGs and $h$RLGs have some blue neighbours, the neighbours may be fainter than the lower luminosity limit ($^{0.1\textrm{\protect\scriptsize KE}}M_{pet}(r) = -20.71$) of our sample.
The pair fraction of $h$REGs is similar to that of $p$REGs, while the pair fraction of $h$BLGs is smaller than that of non-H{\protect\scriptsize II} BLGs (significant up to $2.4\sigma$).
This indicates that the role of close pairs is less important for $h$BLGs than for $h$REGs.

\section{Conclusions}

Here, we summarise the new findings in this paper and their implication on galaxy evolution.

\begin{itemize}
\item[(1)] The morphology-colour class of galaxies strongly depends on the local density. The approximate order of high-density preference is REGs -- RLGs -- BEGs -- BLGs.
\end{itemize}

\begin{itemize}
\item[(2)] In low-density environments, the fraction of AGN host galaxies increases as the local density increases, but it decreases in high-density environments that approximately correspond to cluster environments. This indicates that high-density environments (like cluster environments) suppress AGN activity.
\end{itemize}

\begin{itemize}
\item[(3)] The local luminosity density distribution of $h$REGs is almost the same as that of $p$REGs. It is an open question why the AGN-suppressing mechanisms at high-density environments do not suppress the star formation in $h$REGs.
\end{itemize}

\begin{itemize}
\item[(4)] The non-AGN BEGs prefer intermediate-density environments, while the AGN BEGs prefer low-density environments like BLGs, although with low 
significance. In the context of early-type formation, such preference for low-density environments of BEGs seems to reflect the later formation of early-type galaxies at lower-density environments.
\end{itemize}

\begin{itemize}
\item[(5)] $p$RLGs that have properties very similar to those of REGs (Paper I and II) have environments similar to those of $p$REGs, which confirms the idea that $p$RLGs are intermediate objects between early-type galaxies and late-type galaxies as suggested in Paper I and II.
\end{itemize}

\begin{itemize}
\item[(6)] The pair fraction of $h$REGs does not show statistically significant difference from that of $p$REGs ($1.4\sigma$ significance), while the pair fraction of $h$BLGs is smaller than that of non-H{\protect\scriptsize II} BLGs (significant up to $2.4\sigma$), indicating that the close pair interaction is less important for the star formation of BLGs compared to that of REGs.
\end{itemize}

\begin{itemize}
\item[(7)] Red galaxies show a single red peak in the distribution of the brightest neighbour colour, whereas $h$BLGs and $s$BLGs show double (red + blue) peaks. The brightest neighbours of $s$BLGs tend to be blue, while those of $l$BLGs tend to be red, which shows the possible relationship between AGN types and close pair interactions.
\end{itemize}

\begin{itemize}
\item[(8)] Some ROSAT-detected passive galaxies in high-density environments may be affected by galaxy cluster environments.
However, about 28 per cent (15 of 54) of the ROSAT-detected passive galaxies reside in low-density environments, which are classified as X-ray Bright Optically Normal Galaxies (XBONGs).
\end{itemize}

\section*{Acknowledgements}

JHL appreciates the support and advice of Dr. Eon-Chang Sung.
This work was supported in part by a grant (R01-2007-000-20336-0) from the Basic Research Program of the Korea Science and Engineering Foundation (KOSEF).
CBP and YYC acknowledge the support of the KOSEF through the Astrophysical Research Centre for the Structure and Evolution of the Cosmos (ARCSEC).
Funding for the SDSS and SDSS-II has been provided by the Alfred P. Sloan Foundation, 
the Participating Institutions, the National Science Foundation, 
the US Department of Energy, the National Aeronautics and Space Administration, 
the Japanese Monbukagakusho, the Max Planck Society, and the Higher Education Funding Council for England.
The SDSS Web site is http://www.sdss.org/.
The SDSS is managed by the Astrophysical Research Consortium for the Participating Institutions. 
The Participating Institutions are the American Museum of Natural History, 
Astrophysical Institute Potsdam, the University of Basel, the University of Cambridge, 
Case Western Reserve University, the University of Chicago, Drexel University, Fermilab, 
the Institute for Advanced Study, the Japan Participation Group, Johns Hopkins University, 
the Joint Institute for Nuclear Astrophysics, the Kavli Institute for Particle Astrophysics and Cosmology, 
the Korean Scientist Group, the Chinese Acadeour of Sciences (LAMOST), Los Alamos National Laboratory, 
the Max-Planck-Institute for Astronomy (MPIA), the Max Planck Institute for Astrophysics (MPA), 
New Mexico State University, Ohio State University, the University of Pittsburgh, the University of Portsmouth, 
Princeton University, the US Naval Observatory, and the University of Washington.

\begin{appendix}

\section{Local Density Estimation}\label{denpars} 

\begin{figure}
\includegraphics[width=84mm]{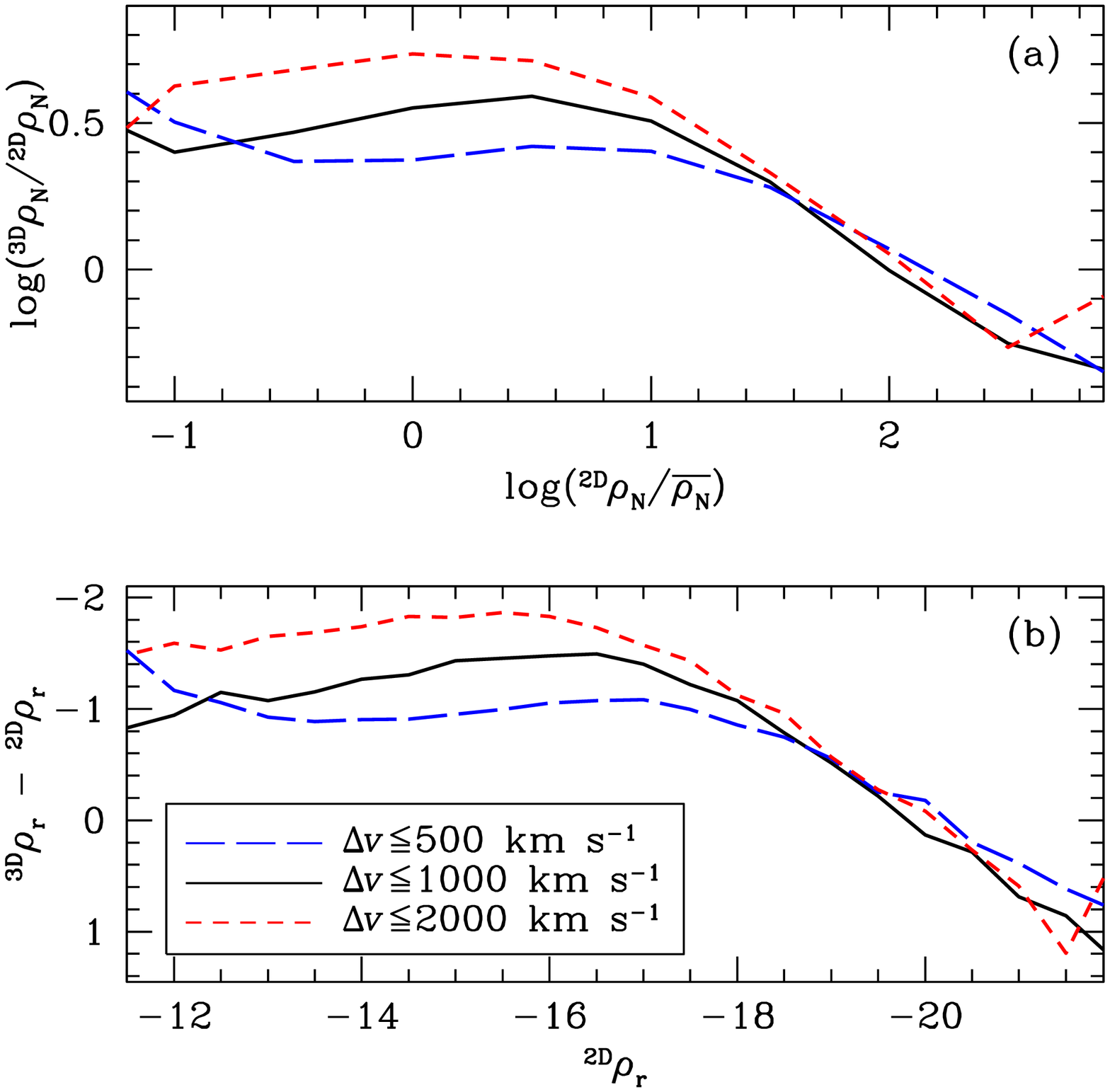}
\caption{ The difference between the 2D and 3D local density estimates as a function of the 2D local density: (a) local number density and (b) local luminosity density. The different line type indicates the different redshift interval in estimating the 2D local density: $\Delta v=$ $\pm500$ km s$^{-1}$ (long-dashed line), $\pm1000$ km s$^{-1}$ (solid line) and $\pm2000$ km s$^{-1}$ (short-dashed line).}
\label{2d3d}
\end{figure}

\subsection{2D and 3D Estimation}\label{A2d3d}

Since the location of a galaxy in the Universe is described in a three-axis frame at given time, we should estimate the number density of galaxies in the 3D space, in principle.
Practically, however, the distance estimation for distant galaxies has very large uncertainty compared to the determination of the 2D coordinates on the plane perpendicular to the line-of-sight, due to the intrinsic limit of the distance estimation.
The distances to galaxies are typically measured by means of determining the recession velocities of galaxies caused by the cosmological expansion. However, it has a problem that the recession velocity of a galaxy is given by both the cosmological expansion and the peculiar motion of that galaxy.

The velocity dispersions of typical rich galaxy clusters range from 500 km s$^{-1}$ to 2000 km s$^{-1}$ \citep{tea90}. Suppose that a galaxy cluster at z $=0.1$ has velocity dispersion of 1000 km s$^{-1}$. In that cluster, 32 per cent of its member galaxies have relative velocities larger than 1000 km s$^{-1}$, compared to the dynamical centre of the cluster, which results in redshift difference larger than 0.003.
If we estimate the distance to the member galaxies with ignoring the peculiar velocity effect, the 0.003 redshift difference is interpreted as the distance difference of 13 Mpc. Since the typical diameter scale of galaxy clusters is at most a few Mpc, this causes significant underestimation of the 3D local number density in rich galaxy clusters.
On the other hand, since such a peculiar velocity effect is very small in the low density fields, the 3D local density is more accurate than the 2D local density in low-density environments.

Fig.~\ref{2d3d} compares between the 2D and 3D local density indicators, showing a clear trend that the 3D local density is smaller than the 2D local density at high-density environments, while the 3D local density is larger than the 2D local density at low-density environments. The difference between the 2D and 3D densities tends to increase as the redshift interval in estimating the 2D local density increases.

\begin{figure}
\includegraphics[width=84mm]{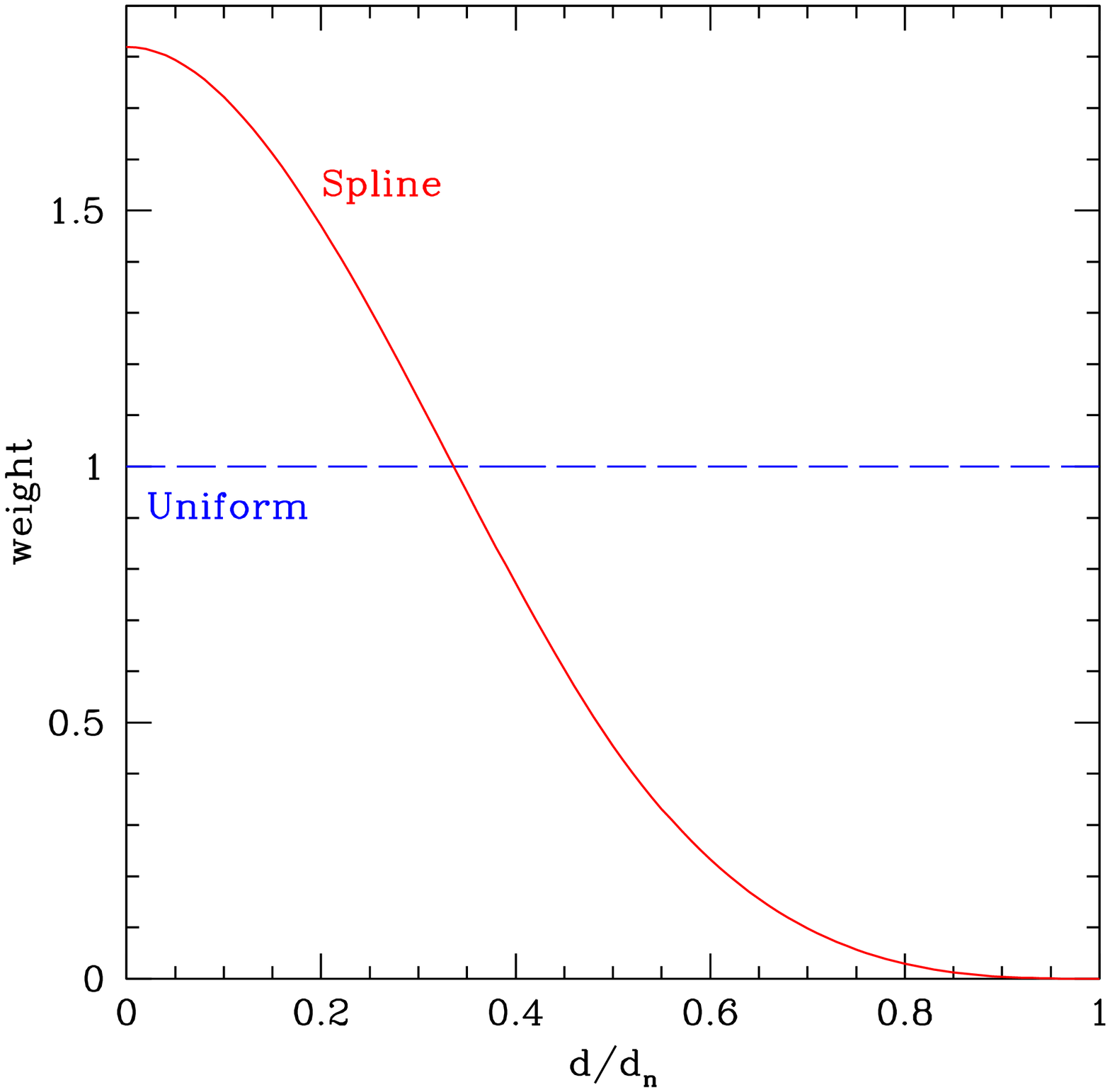}
\caption{ Smoothing kernels for local density estimation as a function of $d/d_n$: uniform (short-dashed) and spline (solid). $d$ is the distance from the target object, and $d_{n}$ is the kernel size. }
\label{kernel}
\end{figure}

\subsection{Smoothing Kernel}\label{Akernel}

We adopted a spline smoothing kernel when estimating the local density, because it is centrally weighted unlike the uniform kernel and because it has a finite size unlike the Gaussian kernel. The basic functional form of the spline kernel is described well in \citet{cab08}. In this paper, the spline kernel is defined as:
\begin{equation}
f_{sp}(d/d_n) = A \left\{ \begin{array}{ll} 1-6(d/d_n)^2+6(d/d_n)^3 & 0 \le d/d_n \le 0.5 \\
2(1-d/d_n)^3 & 0.5 < d/d_n \le 1 \\
0 & d/d_n > 1 \\
 \end{array} \right.
\end{equation}
where $d$ is the distance from the target object and $d_{n}$ is the kernel size. $A$ is a constant, which has the value of $40/7$ and $32/3$ in the 2D and 3D local density estimation, respectively. Fig.~\ref{kernel} shows the shape of the spline kernel used in this paper.

\begin{figure*}
\includegraphics[width=168mm]{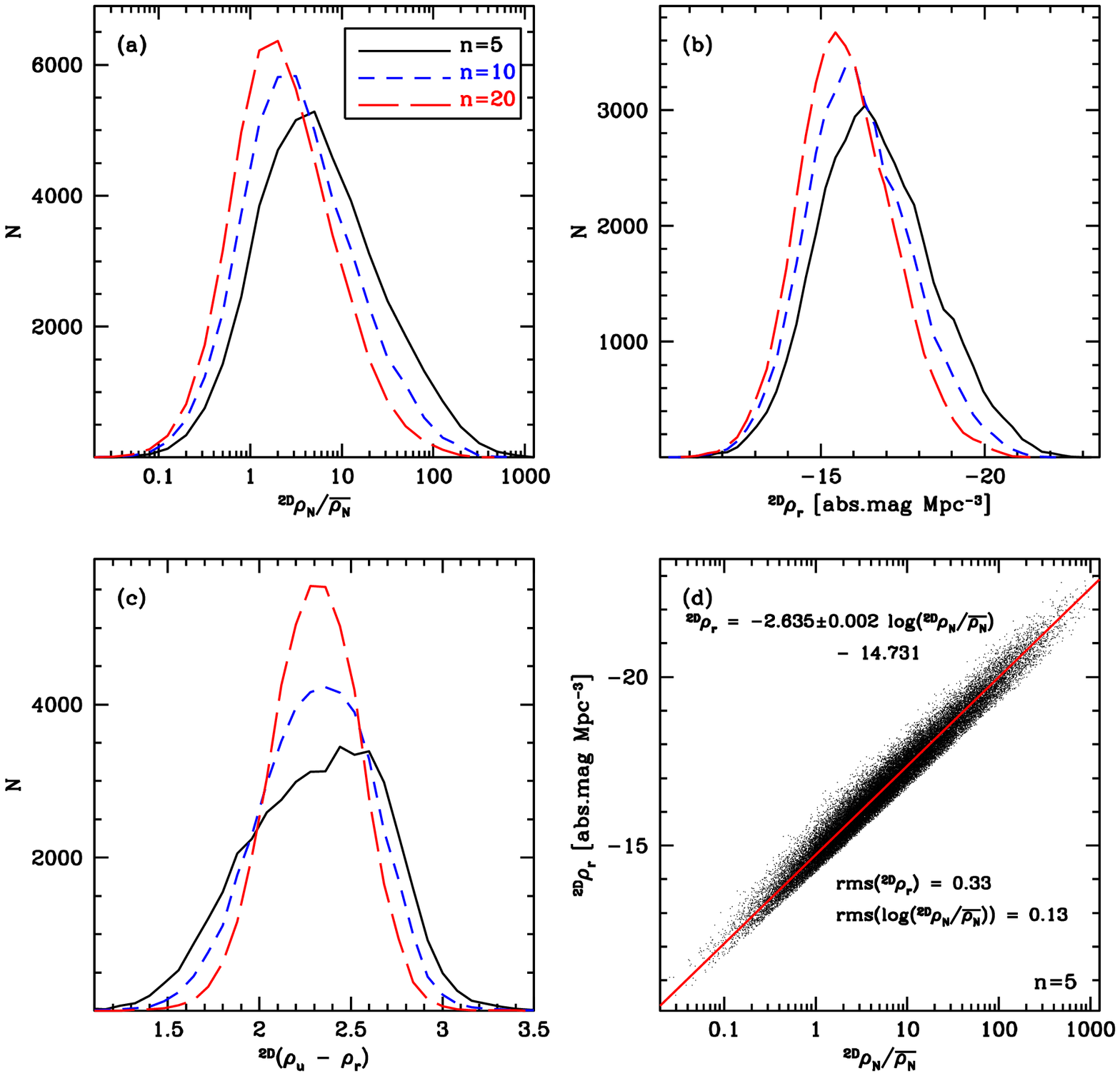}
\caption{ (a) The distribution of the 2D local number density ($^{\textrm{\protect\tiny{2D}}}{\rho}_N$) divided by the mean number density ($\bar{\rho}_N$) for the SDSS galaxies in the volume-limited sample. The solid line, short-dashed line and long-dashed line show the results using $d_5$, $d_{10}$ and $d_{20}$, respectively. (b) The 2D local luminosity density ($^{\textrm{\protect\tiny{2D}}}{\rho}_r$) distribution in the $r$ band. (c) Local colour  ($^{\textrm{\protect\tiny{2D}}}(\rho_u-\rho_r)$) distribution. (d) The local number density -- local luminosity density relation, using $d_5$. The line shows the least-squares fit using the ordinary least-squares bisector method \citep{iso90}. }
\label{basic}
\end{figure*}

\begin{figure*}
\includegraphics[width=168mm]{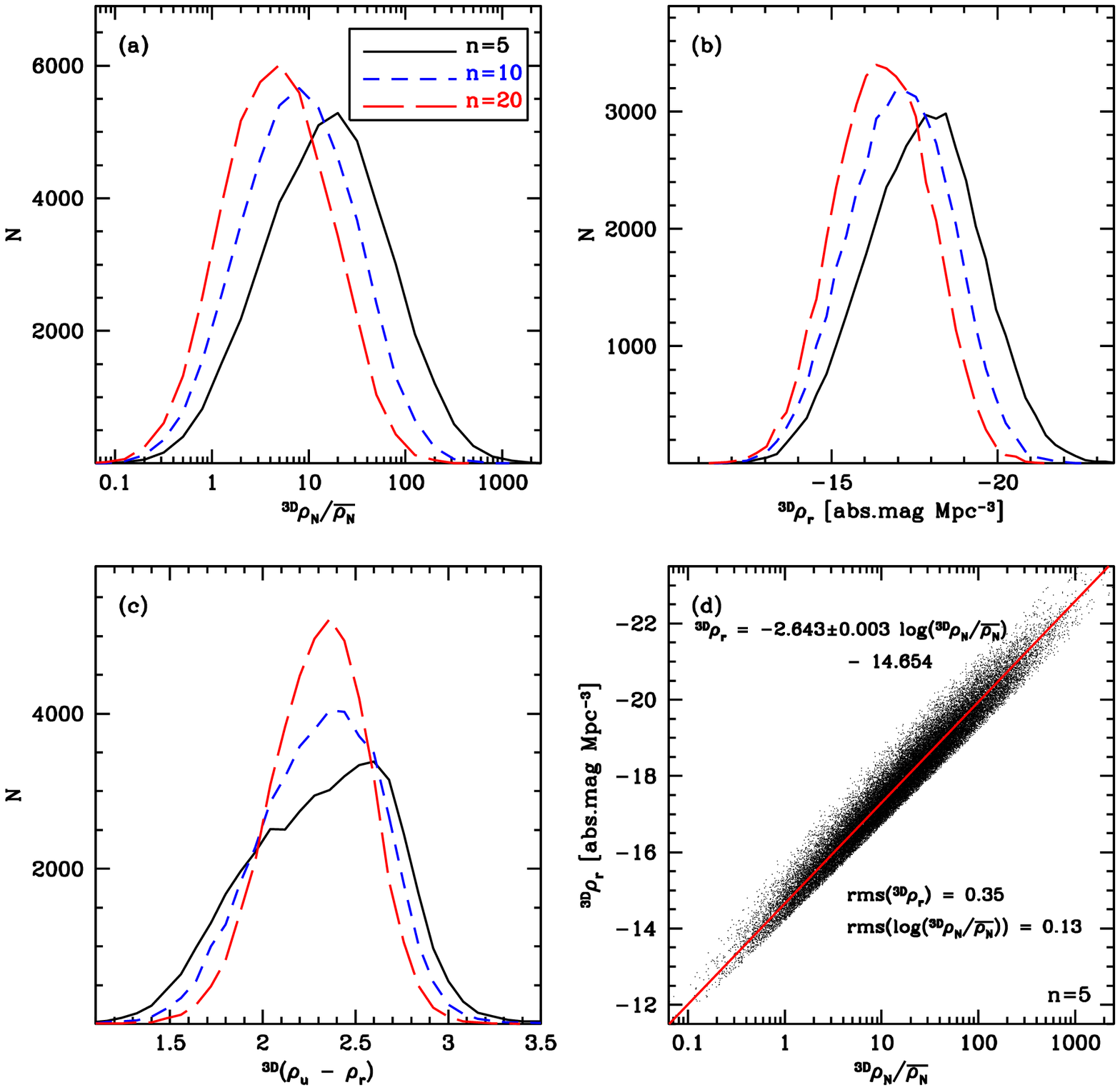}
\caption{ The same as Fig.~\ref{basic}, but for the 3D local density indicators. }
\label{basic2}
\end{figure*}

\subsection{Local Density Indicators}\label{defld}

As a local density indicator, the local galaxy number density is quite useful but has some weakness at the same time.
That is, in the local number density estimation, both five faint blue galaxies and five bright red galaxies are commonly counted as \emph{five galaxies}, which is missing the luminosity and colour information of the counted galaxies.
The luminosity dispersion problem can be minimised by limiting the luminosity of the counted galaxies to a narrow range \citep{par07}. However, another way to reflect both the luminosity and colour information of the galaxies in the local environments is to estimate the local luminosity density and the local colour. It should be reminded that the local luminosity density is not the same as the local mass density \citep{par08,par09a}, because galaxies with the same mass may have quite different luminosity according to their mass-to-light ratio. Thus, both the local number density and the local luminosity density need to be compared for more reliable environmental analysis.

Fig.~\ref{basic}a shows the distribution of the local galaxy number density ($\rho_N$) divided by the mean number density in our sample, $\bar{\rho}_{N}=2.469\times10^{-3}$ Mpc$^{-3}$. Fig.~\ref{basic}b and Fig.~\ref{basic}c are the same as Fig.~\ref{basic}a, but for the local luminosity density ($\rho_r$) and the local colour ($\rho_u-\rho_r$), respectively.
When the larger $n$ value of the kernel size $d_n$ is adopted, the distributions of $\rho_N/\bar{\rho}_{N}$, $\rho_{r}$ and $\rho_u-\rho_r$ become shallower, showing that a large kernel tends to smooth out the variations of small-scale environments.
Since the main interest of this paper is the local environments that affect directly the properties of galaxies rather than the large-scale structure, we focus on the small-scale environments by adopting the $d_5$ kernel.
Fig.~\ref{basic}d and \ref{basic2}d display the local number density versus the local luminosity density in the 2D and 3D estimation, showing tight correlations with the ordinary least-squares bisector \citep{iso90} fit:
\begin{equation}\label{Andld}
^{\textrm{\protect\tiny{2D}}}\rho_r = -2.635\pm0.002\log(^{\textrm{\protect\tiny{2D}}}\rho_N/\bar{\rho}_N)-14.731
\end{equation}
\begin{equation}\label{Andld2}
^{\textrm{\protect\tiny{3D}}}\rho_r = -2.643\pm0.003\log(^{\textrm{\protect\tiny{3D}}}\rho_N/\bar{\rho}_N)-14.654
\end{equation}
with $\rho_r$-direction rms of 0.33 and 0.35 absolute-mag Mpc$^{-3}$, respectively. That is, the local luminosity density reflects well the local number density overall, but there are typically $\rho_r$ dispersions of $\pm0.33$ ($\pm0.35$) mag Mpc$^{-3}$ at fixed $\rho_N$ in the 2D (3D) estimation.

\begin{figure*}
\includegraphics[width=168mm]{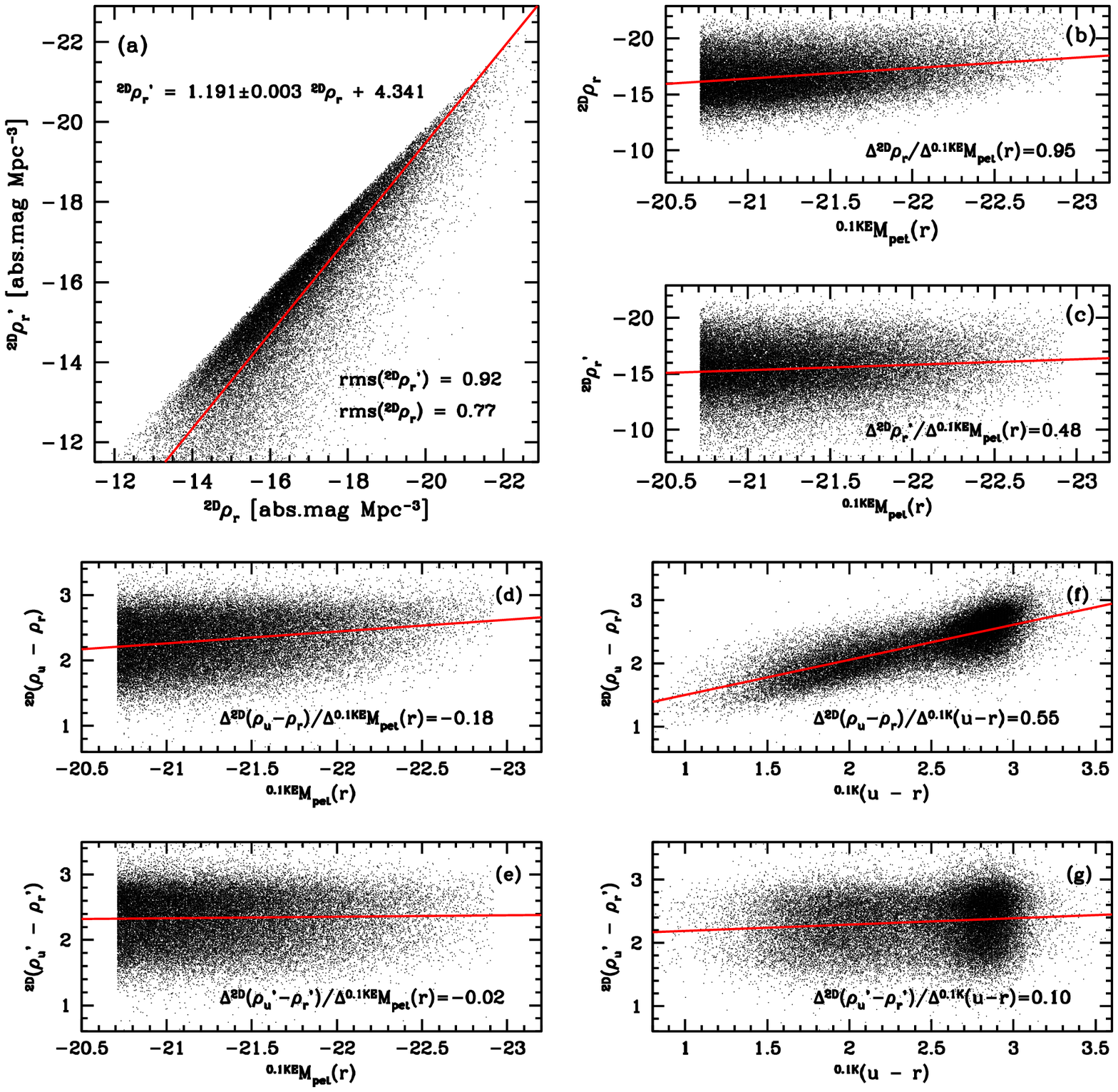}
\caption{ (a) The comparison between the original 2D local luminosity density and the target-excluded 2D local luminosity density. The line shows the ordinary least-squares bisector fit. (b) The dependence of the original 2D local luminosity density on the target luminosity. (c) The dependence of the target-excluded 2D local luminosity density on the target luminosity. (d) The dependence of the original 2D local colour on the target luminosity. (e) The dependence of the target-excluded 2D local colour on the target luminosity. (f) The dependence of the original 2D local colour on the target colour. (g) The dependence of the target-excluded 2D local colour on the target colour. }
\label{prime}
\end{figure*}

\begin{figure*}
\includegraphics[width=168mm]{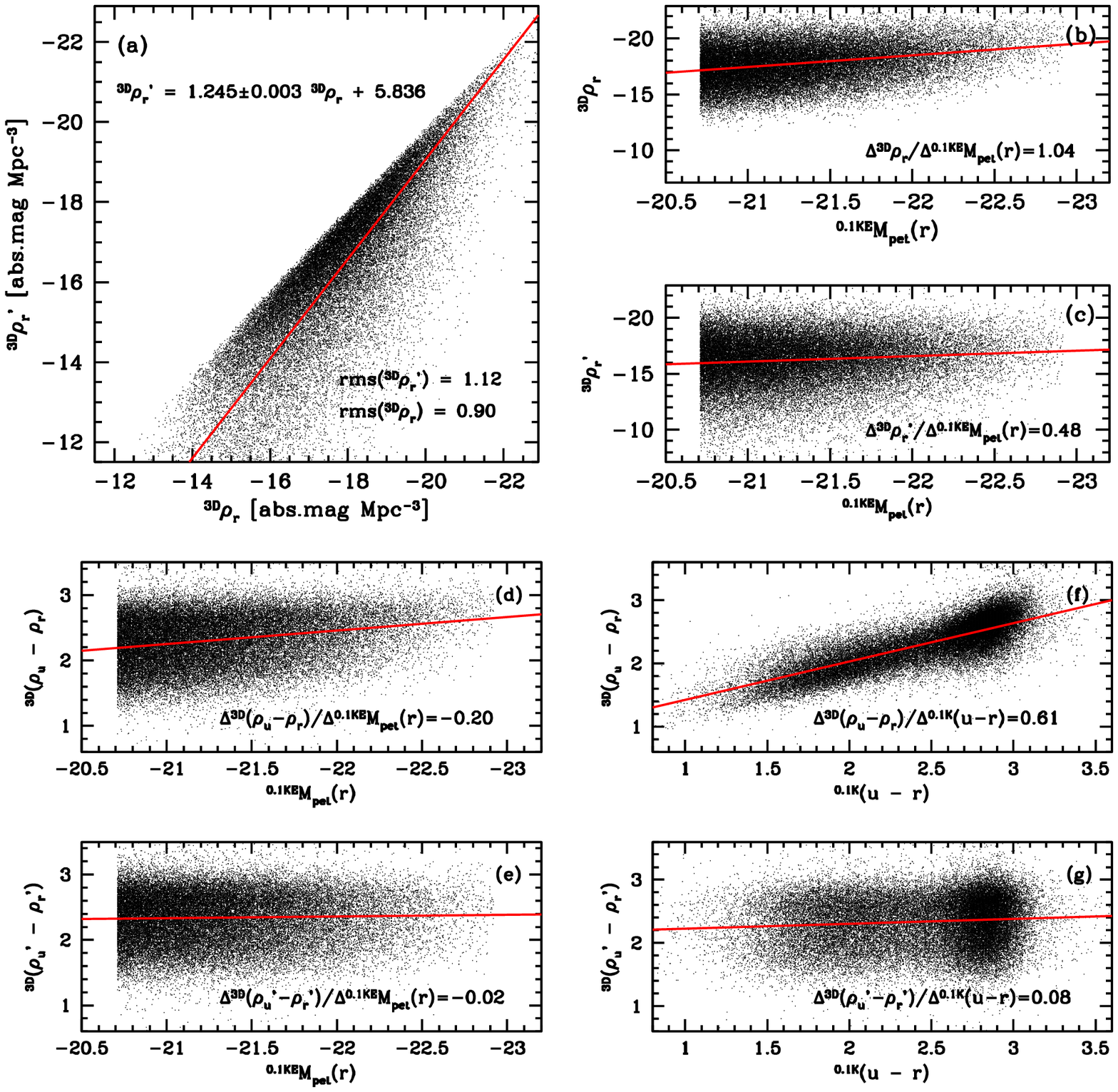}
\caption{ The same as Fig.~\ref{prime}, but for the 3D local density indicators. }
\label{prime2}
\end{figure*}

\subsection{Corrected Environmental Parameters}\label{Acorr}

The local luminosity density and the local colour provide additional information that the local number density does not show. However, there is one issue we should consider: $\rho_r$ and $\rho_u-\rho_r$ include the information of the target galaxy itself. In the local number density, this is not a significant problem, because the target galaxy itself is always counted as simply \emph{one galaxy}, which affects the zeropoint of the environmental parameter but does not cause any systematic distortion. On the other hand, this may cause a significant distortion for the local luminosity density and the local colour, because target galaxies have their own luminosity and colour ranging widely.

Suppose a target galaxy and its several neighbour galaxies and we calculate the local luminosity density and the local colour for the target galaxy. After that, we replace the target galaxy with a new one with quite different luminosity and colour. Then, the new local luminosity density and local colour may be different from the old values. However, the local environment has not changed actually, but it is only the target galaxy that changed.
Therefore, it is reasonable to exclude the information of the target galaxy itself when estimating the local luminosity density and the local colour, to investigate the environmental effects on the target galaxy.
The local number or mass density used by \citet{par07,par08} and \citet{par09a} excluded the contribution of the target galaxy itself.

Fig.~\ref{prime} and \ref{prime2} compare the original environmental parameters ($\rho_r$ and $\rho_u-\rho_r$) and the target-excluded environmental parameters (${\rho_r}'$ and ${\rho_u}'-{\rho_r}'$).
${\rho_r}'$ is systematically fainter than $\rho_r$ (because only four galaxies are used in the luminosity summation, rather than five), but still correlated with $\rho_r$ (Fig.~\ref{prime}a and \ref{prime2}a). In Fig.~\ref{prime}b,d,f and \ref{prime2}b,d,f, it seems that $\rho_r$ and $\rho_u-\rho_r$ strongly depend on the luminosity and colour of the target galaxy. However, such dependency becomes very weak when the $\rho_r$ and $\rho_u-\rho_r$ are replaced by ${\rho_r}'$ and ${\rho_u}'-{\rho_r}'$, the target-excluded environmental parameters, as shown in Fig.~\ref{prime}c,e,g and \ref{prime2}c,e,g.
After excluding the information of the target galaxy, the environmental parameters have correlations with the target luminosity and colour, as follows:
\begin{equation}
^{\textrm{\protect\tiny{2D}}}{\rho_r}' = 0.484\times{^{0.1\textrm{\protect\scriptsize KE}}M_{\textrm{\protect\scriptsize pet}}(r)} - 5.171
\end{equation}
\begin{equation}
^{\textrm{\protect\tiny{2D}}}({\rho_u}'-{\rho_r}') = 0.101\times{^{0.1\textrm{\protect\scriptsize K}}(u-r)} + 2.091
\end{equation}
\begin{equation}
^{\textrm{\protect\tiny{3D}}}{\rho_r}' = 0.480\times{^{0.1\textrm{\protect\scriptsize KE}}M_{\textrm{\protect\scriptsize pet}}(r)} - 6.002
\end{equation}
\begin{equation}
^{\textrm{\protect\tiny{3D}}}({\rho_u}'-{\rho_r}') = 0.077\times{^{0.1\textrm{\protect\scriptsize K}}(u-r)} + 2.151,
\end{equation}
which show relatively weak dependence on the local environments, compared to the target-included local density indicators.

\end{appendix}

\label{lastpage}

\end{document}